\documentclass{jpp}
\usepackage[utf8]{inputenc}
\usepackage[T1]{fontenc}
\usepackage{xcolor, mathtools, graphicx, dcolumn, bm, lipsum, tikz, pgfplots, afterpage, capt-of, pdflscape}
\usepackage[colorlinks=true,citecolor=blue,linkcolor=blue]{hyperref}

\usetikzlibrary{intersections}
\usetikzlibrary{patterns}

\newcommand{\eq}[2]{\begin{equation} \label{eq:#1} #2 \end{equation}}

\newcommand{\bs}[1]{\boldsymbol{#1}}
\def\be{\begin{equation}}
\def\ee{\end{equation}}
\def\ba{\begin{eqnarray}}
\def\ea{\end{eqnarray}}
\newcommand\bx{\boldsymbol{x}}
\newcommand\bu{\boldsymbol{u}}

\newcommand\bB{\boldsymbol{B}}
\newcommand\bb{\boldsymbol{b}}
\newcommand\bE{\boldsymbol{E}}
\def\pdt{\partial_t}

\def\pdy{\partial_y}

\def\div{\bnabla \cdot}
\def\rot{\bnabla \times}
\def\dd{\mathrm{d}}

\shorttitle{Wave turbulence in IEMHD}
\shortauthor{V. David \& S. Galtier}

\title{Wave Turbulence in Inertial Electron Magnetohydrodynamics}

\author{Vincent David \aff{1,2}
  \corresp{\email{vincent.david@lpp.polytechnique.fr}}
 \and Sébastien Galtier \aff{1,2,3}}

\affiliation{\aff{1}Laboratoire de Physique des Plasmas, \'Ecole polytechnique, F-91128 Palaiseau Cedex, France
\aff{2}Universit\'e Paris-Saclay, IPP, CNRS, Observatoire Paris-Meudon, France
\aff{3}Institut universitaire de France}

\begin{document}

\maketitle

\begin{abstract}
A wave turbulence theory is developed for inertial electron magnetohydrodynamics (IEMHD) in the presence of a relatively strong and uniform external magnetic field $\bs{B_0} = B_0 \hat{\bs{e}}_\|$. This regime is relevant for scales smaller than the electron inertial length $d_e$. We derive the kinetic equations that describe the three-wave interactions between inertial whistler or kinetic Alfv\'en waves. We show that for both invariants, energy and momentum, the transfer is anisotropic (axisymmetric) with a direct cascade mainly in the direction perpendicular ($\perp$) to $\bs{B_0}$. The exact stationary solutions (Kolmogorov-Zakharov spectra) are obtained for which we prove the locality. We also found the Kolmogorov constant $C_K \simeq 8.474$. 
In the simplest case, the study reveals an energy spectrum in $k_\perp^{-5/2} k_\|^{-1/2}$ and a momentum spectrum enslaved to the energy dynamics in $k_\perp^{-3/2} k_\|^{-1/2}$. These solutions correspond to a magnetic energy spectrum $\sim k_\perp^{-9/2}$, which is steeper than the EMHD prediction made for scales larger than $d_e$. We conclude with a discussion on the application of the theory to space plasmas. 
\end{abstract}

\section{Introduction}\label{sec:Introduction}

There are many ways to investigate the problem of energy transfer through the different scales in a turbulent plasma, but one of the most rigorous is the theory of wave turbulence. This theory is limited to systems composed of a sea of weakly interacting waves. Since the nonlinearities are weak, it is possible to describe how the dynamics of the system develops in a time asymptotically long compared to the period of the waves considered \citep{Nazarenko_2011}. The importance of the wave turbulence theory is, first, the fact that a natural closure can be achieved with a uniformity of the asymptotic development \citep{Benney1966,Benney_1969} and, second, the possibility to derive exact solutions (Kolmogorov-Zakharov spectra) of the wave kinetic equations \citep{Zakharov_1992}. 
The wave turbulence regime is a highly studied subject in physics, both theoretically and experimentally. Examples are provided e.g. in hydrodynamics with surface waves \citep{Zakharov_1967,Falcon_2022}, internal gravity waves \citep{Caillol_2000,Dematteis_2021} and inertial waves \citep{Galtier_2003a,Yarom_2014,Monsalve_2020}; in plasma physics with MHD \citep{Galtier_2000,Kuznetsov_2001,Meyrand_2015}, Hall-MHD \citep{Galtier_2006,Meyrand_2018} and rotating MHD \citep{Galtier_2014}; in acoustic waves \citep{Zakharov_1970,Newell_1971,Lvov_1997}, optical waves \citep{Dyachenko_1992}, elastic waves \citep{During_2006,Hassaini_2019}, Kelvin waves \citep{Laurie_2010}, in Bose-Einstein condensates \citep{Nazarenko_2006} and even with gravitational waves \citep{Galtier_2017,Galtier2021}. 

Here, we are interested in the plasma dynamics at scales where the electron inertia plays a non-negligible role. In our approach, the mass difference between ions and electrons is such that the ions will be considered static to form a neutralizing background. Therefore, at the time scale of interest, only the electron dynamics is relevant. This is the domain of EMHD and IEMHD which describe, respectively, the scales $\ell \gg d_e$ and $d_e \gg \ell \gg r_e$, where $r_e$ is the electron Larmor radius. Our study will focus on the latter case. Although it is difficult for current spacecraft to measure the plasma dynamics corresponding to the electron inertia scales, it is interesting to see what the theoretical description can predict. 
The EMHD and IEMHD approximations are widely used models to study, for example, magnetic reconnection or space plasma turbulence \citep{Bulanov1992,Biskamp1996,Biskamp1999,Dastgeer2000b,Dastgeer2000,Cho2004,Cho2011,Kim2015}. More information is given in \cite{Milanese_2020} where an exhaustive list of plasmas driven by the IEMHD model is given with the parameter regimes.
In this paper, we present the theory of wave turbulence for IEMHD in the presence of a relatively strong and uniform external magnetic field $\bs{B_0}$. The equivalent theory for EMHD has already been published \citep{Galtier_2003b} but not yet for IEMHD. Strong IEMHD turbulence has recently received new attention with the study of the weakly compressible case \citep{Chen_2017,Roytershteyn_2019}. The objective was to study the nature of plasma turbulence in the Earth's magnetosheath. The main prediction, phenomenological in nature, is a magnetic spectrum in $k_\perp^{-11/3}$ (see also \cite{Meyrand_2010}) which is less steep than the prediction we will derive in this paper. In the meantime, a rigorous derivation (using systematic asymptotic expansions) based on a more general model including electron inertia and finite Larmor radius corrections has been proposed \citep{Passot_2017,Passot_2019}. This more general approach allows the study of several different limits, and to recover in particular the model discussed previously \citep{Chen_2017}. In fact, this weakly compressible IEMHD equations have the same mathematical structure as the incompressible case when the ion $\beta_i$ (the ratio between ion thermal pressure and magnetic pressure) is moderately small. Therefore, the physics of wave turbulence that we will describe in this paper has a broader impact than strictly speaking the incompressible case and can be applied for both inertial whistler waves (IWW) and inertial kinetic Alfv\'en waves (IKAW). 
A similar situation exists for scales larger than $d_e$: in the presence of a strong $\bs{B_0}$, the equations describing the nonlinear dynamics of kinetic Alfv\'en waves and whistler waves have exactly the same mathematical form, which means that the physics of wave turbulence is similar for both problems \citep{Galtier2015}. 
Although a fully kinetic approach is {\it a priori} required to describe plasma dynamics at electron inertial scales, all of these reduced fluid models can provide interesting insight when considering small fluctuations around a Maxwellian equilibrium state. In this paper, we follow this precept and apply the powerful tool of wave turbulence to extract new properties useful for a better understanding of space plasmas. 

The structure of the article is as follows. In Section \ref{sec:Derivation}, we propose a quick (and therefore simplified) derivation of the system of equations that we will use for the theory of wave turbulence. In Section \ref{sec:Triadic}, we introduce the canonical variables and derive the dynamical equation describing the wave amplitude variation. In Section \ref{sec:Phenomenology}, a phenomenology of wave turbulence is developed to get a simple heuristic explanation to the solutions (Kolmogorov-Zakharov spectra) derived later. In Section \ref{sec:Kinetic}, we derive the wave  kinetic equations from which we show the detailed conservation of invariants. In section \ref{sec:Spectra}, we obtain the exact stationary solutions in the anisotropic limit $k_\| \ll k_\perp$ and the locality of these solutions is proved. In Section \ref{sec:SuperLocal}, we consider the limit of super-local interactions and derive the associated nonlinear diffusion equation for the energy. In Section \ref{sec:Flux}, we compute the sign of the energy flux which gives the direction of the cascade, and find the Kolmogorov constant. We conclude in Section \ref{sec:Discussion} with a discussion of possible applications of our results, and ways in which these results can be extended.

\section{Inertial electron magnetohydrodynamics}\label{sec:Derivation}

The goal of this section is to quickly derive in a simplified way the set of equations describing the dynamics of non-relativistic electrons at inertial scales in a fully ionized plasma. For this reason, the assumption of incompressibility will be used. A complete derivation is found in \cite{Chen_2017} and in \cite{Passot_2017}.

\subsection{Governing equations}

The basic fluid equations governing the electron dynamics in an incompressible (dissipationless) plasma are
\begin{eqnarray}
\label{eq:dudt}
\pdt \bu_e + \left(\bu_e \cdot \bnabla\right) \bu_e &=& -\frac{1}{\rho_e} \bnabla P_e - \frac{q_e}{m_e} \left( \bu_e \times \bB + \bs{E} \right), \\
\label{eq:dBdt}
\pdt \bB &=& - \bnabla \times \bE, \\
\label{eq:MaxwellAmpere}
\rot \bB &=& \mu_0 \bs{J}, \\
\label{eq:divu}
\div \bu_e &=& 0, \\
\label{eq:divB}
\div \bB &=& 0 ,
\end{eqnarray}
where $\bu_e(\bx,t)$ is the electron velocity, $\rho_e(\bx,t)= m_e n_0$ the constant electron mass density with $m_e$ the electron mass and $n_0$ the density, $P_e(\bx,t)$ the electron pressure, $q_e>0$ the modulus of the electron charge, $\bB(\bx,t)$ the magnetic field, $\bs{E}(\bx,t)$ the electric field, $\boldsymbol{J}(\bx,t) = n_0 q_e \left(\bu_i - \bu_e \right)$ the electric current and $\bu_i(\bx,t)$ the ion velocity (assumed to be zero). 
Normalizing the magnetic field to the (electron) Alfv\'en velocity and then taking the rotational of equation (\ref{eq:dudt}) combined with the Maxwell-Faraday law (\ref{eq:dBdt}), one obtains
\eq{}{
\pdt \left( d_e^2 \nabla^2 -1\right) \bb + \left( \bu_e \cdot \bnabla \right) (d_e^2 \nabla^2 -1)\bb
= \left( d_e^2 \nabla^2 - 1 \right) \bb \cdot \bnabla \bu_e,
}
where $d_e = \sqrt{m_e/(n_0 q_e^2 \mu_0)}$ is the electron inertial length. Now, we introduce a relatively strong and uniform (normalized) magnetic field $\bb_0 = b_0 \hat{\bs{e}}_\|$ that defines the parallel direction. In the limit of IEMHD, the spatial variations of $\bb$ are done on a characteristic length $L \ll d_e$ and mainly in the plane perpendicular to $\hat{\bs{e}}_\|$. Thus at the leading order, we have
\eq{IKAW}
{
\left( \pdt + {\bu_e}_\perp \cdot \bnabla_\perp \right)d_e^2 \nabla^2_\perp  \bb  = d_e^2 (\nabla^2_\perp \bb_\perp \cdot \bnabla_\perp) \bu_e - (\bb_0 \cdot \bnabla) \bu_e,
}
and also $\boldsymbol{J} = - n_0 q_e \bu_e$, which can be written $d_e \bs{j} = - \bu_e$
with the normalized electric current $\bs{j} \equiv \bnabla \times \bb$. 
The magnetic field having a zero divergence, we define $\bb \equiv \bb_0 - \bnabla \times \left( g \boldsymbol{e}_x + \psi \boldsymbol{e}_z \right)$ where $ \hat{\boldsymbol{e}}_z$ is a unit vector (hereafter, we will assume $\hat{\bs{e}}_z = \hat{\bs{e}}_\|$ which is valid at leading order for a relatively strong uniform magnetic field $\bb_0$), $\psi(\bx,t)$ a stream function and $g(\bx,t)$ a function satisfying the relation $\pdy g \equiv b_\| $. We obtain the relation
\eq{neweq}{
\nabla^2_\perp  \bb  = \left( \hat{\boldsymbol{e}}_\| \times \bnabla_\perp \right) (\nabla^2_\perp \psi) + \nabla^2_\perp b_\| \hat{\boldsymbol{e}}_\|,
}
where, hereafter, the $z$-derivative is assumed to be negligible compared to the perpendicular derivative. Replacing $\bb$ by its expression, the electron velocity can be expressed as a function of the magnetic field components
\eq{defVelocity}{
\bu_e = d_e \left( \hat{\boldsymbol{e}}_\| \times \bnabla_\perp b_\| - \nabla_\perp^2 \psi \hat{\boldsymbol{e}}_\| \right).
}
Projecting equation (\ref{eq:IKAW}) in the perpendicular plane to $ \hat{\boldsymbol{e}}_\|$, we find
\eq{}{
\begin{split}
    & \left( \hat{\boldsymbol{e}}_\| \times \bnabla_\perp \right) \left[\pdt \left( d_e^2 \nabla_\perp^2 \psi \right) \right] + d_e^2 \left( {\bu_e}_\perp \cdot \bnabla_\perp \right) \left[ \left( \hat{\boldsymbol{e}}_\| \times \bnabla_\perp \right) \nabla^2_\perp \psi \right] \\
    &= d_e^3 \left( \nabla^2_\perp \bb \cdot \bnabla_\perp \right) \left( \hat{\boldsymbol{e}}_\| \times \bnabla_\perp b_\| \right) - d_e b_0 \p_\| \left( \hat{\boldsymbol{e}}_\| \times \bnabla_\perp \right) b_\| .
\end{split}
}
The non-trivial relation
\eq{}{
\begin{split}
&\left( {\bu_e}_\perp \cdot \bnabla_\perp \right) \left[ \left( \hat{\boldsymbol{e}}_\| \times \bnabla_\perp \right) \nabla^2_\perp \psi \right] = \\
&
\left(\hat{\boldsymbol{e}}_\| \times \bnabla_\perp \right) \left[ \left({\bu_e}_\perp \cdot \bnabla_\perp \right) \nabla^2_\perp \psi \right] + d_e \left( \nabla^2_\perp \bb \cdot \bnabla_\perp \right) \left( \hat{\boldsymbol{e}}_\| \times \bnabla_\perp b_\| \right),
\end{split}
}
allows to simplify the previous equation and, by expressing $\bu_e$ as a function of $\psi$, we obtain after some algebraic manipulations
\eq{IKAWperp}{
\pdt \left( \nabla_\perp^2 \psi \right) + d_e \left[ \left( \hat{\boldsymbol{e}}_\| \times \bnabla_\perp b_\| \right) \cdot \bnabla_\perp \right] \nabla_\perp^2 \psi  = - \Omega_e \p_\| b_\|,
}
with $\Omega_e \equiv b_0 / d_e $ the cyclotron frequency of electrons (note that here, $\Omega_e$ is constant due to the assumption of incompressibility).

Now, a projection of (\ref{eq:IKAW}) in the $ \hat{\boldsymbol{e}}_\|$ direction gives directly
\eq{}{
\pdt \left( d_e^2 \nabla_\perp^2 b_\| \right) + d_e^2 \left( {\bu_e}_\perp \cdot \bnabla_\perp \right) \nabla^2_\perp b_\|  = - d_e^3 \left( \nabla^2_\perp \bb \cdot \bnabla_\perp \right)\nabla_\perp^2 \psi + d_e b_0 \p_\| \left( \nabla_\perp^2 \psi \right).
}
It is straightforward to show that the first term of the right-hand side is exactly zero. Then, by expressing $\bb$ and $\bu_e$ as functions of $\psi$ and $b_\|$, we obtain
\eq{IKAWpara}{
\pdt \left( \nabla_\perp^2 b_\| \right) + d_e \left[ \left( \hat{\boldsymbol{e}}_\| \times \bnabla_\perp b_\| \right) \cdot \bnabla_\perp \right] \nabla_\perp^2 b_\| =  \Omega_e \p_\| \left( \nabla_\perp^2 \psi \right)
.}

Equations (\ref{eq:IKAWperp}) and (\ref{eq:IKAWpara}) describe the dynamics of electrons at inertial scales. They have been derived in a more general framework and using kinetic arguments by \cite{Chen_2017} and \cite{Passot_2017}. Here, we have used the incompressibility condition to propose a (less accurate but more) fast derivation of a system that a priori describes only IWW. However, it is interesting to note that at inertial electron scales: (i) IKAW and IWW can have the same dispersion relation and the only difference is that the transition to the inertial regime occurs at $k_\perp^2 d_e^2 \simeq 1$ for IWW rather than $k_\perp^2 d_e^2 \simeq 1 + 2/ \beta_i$ for IKAW; (ii) the nonlinear equations governing the dynamics of IKAW and IWW are mathematically similar (up to a change of variable from $b_z$ to $\rho_e$ \citep{Chen_2017,Passot_2017}), which means that the physics of wave turbulence developed in this paper applies to both waves. 
A similar situation is found at scales larger than $d_e$: in the presence of a strong $\bs{B_0}$, the equations describing the nonlinear dynamics of kinetic Alfv\'en waves and whistler waves have exactly the same mathematical form, which means that the physics of wave turbulence is similar for both problems \citep{Galtier2015}. 


\subsection{Three-dimensional quadratic invariants} \label{subsec:invariants}

In the absence of forcing and dissipation, the system (\ref{eq:IKAWperp})--(\ref{eq:IKAWpara}) has two quadratic invariants. The first  invariant is the energy which is written at the leading order
\eq{totalEnergy}{
E = d_e^2 \left\langle \bs{j}^2 \right\rangle = E_\perp + E_\| = d_e^2 \left\langle \left(\nabla_\perp b_\| \right)^2 + \left(\nabla_\perp^2 \psi \right)^2 \right\rangle,
}
where $\langle \rangle$ is a spatial average or, equivalently by ergodicity, an ensemble average. $E$ can also be interpreted as the kinetic energy of electrons. As shown in Appendix \ref{appendix:Conservation}, both $E_\perp$ and $E_\|$ are separately conserved at the nonlinear level, however, energy is exchanged between the two at the linear level, thanks to the presence of waves. This definition of energy is valid for both IWW and for IKAW in the limit of small $\beta_i$. 

The second quadratic invariant is the momentum that can be written at the leading order
\eq{totalHelicity}{
H = d_e^2 \left\langle (\nabla_\perp^2 \psi) (\nabla_\perp^2 b_\|) \right\rangle.
}
$H$ can be interpreted as the kinetic helicity of electrons. Unlike energy, the momentum is not positive defined. As we will see later, the wave kinetic equations conserve these two invariants on the resonant manifold.

\subsection{Dispersion relation}

In the linear regime, the Fourier transform of equations  (\ref{eq:IKAWperp}) and (\ref{eq:IKAWpara}) gives
\begin{eqnarray}
\pdt \psi_k  &=& i \Omega_e k_\| k_\perp^{-2} b_k, \\
\pdt b_k &=&  i \Omega_e k_\| \psi_k,
\end{eqnarray}
where the Fourier transform used is
\eq{}{
\psi(\bs{k},t) \equiv \psi_k = \int_{\mathbb{R}^3} \psi(\bx,t) \mathrm{e}^{-i \bs{k} \cdot \bs{x}} \dd \bs{x}.
}
Hereafter, we use the notation $b_k \equiv {b_\|}_k$. If the wavevector $\bs{k}$ is decomposed as $\bs{k} = k_\perp \hat{\bs{e}}_\perp + k_\| \hat{\bs{e}}_\|$, then the linear dispersion relation reads
\eq{dispersionRelation}{
\left( \frac{\omega_k}{\Omega_e} \right)^2 = \left( \frac{k_\|}{k_\perp} \right)^2 .
}
One can find the following solutions to the linear IEMHD equations in Fourier space
 \begin{eqnarray}
\psi_k(k_\perp,t) &=& f \left( k_\perp \right) \cos\left(\omega_k t \right) + i g \left( k_\perp \right) k_\perp^{-1} \sin \left(\omega_k t \right), \\
b_k(k_\perp,t) &=& g \left( k_\perp \right) \cos\left(\omega_k t \right) + i f \left( k_\perp \right) k_\perp \sin \left(\omega_k t \right),
\end{eqnarray}
with $f$ and $g$ two arbitrary functions.

\section{Wave amplitude equation}\label{sec:Triadic}

In Fourier space, IEMHD equations (\ref{eq:IKAWperp}) and (\ref{eq:IKAWpara}) become
\begin{eqnarray}
\label{eq:ikawPsiFourier}
k_\perp^2 \pdt \psi_k - i \Omega_e k_\| b_k &=& d_e\int_{\mathbb{R}^6} \hat{\bs{e}}_\| \cdot \left(\bs{p}_\perp \times \bs{q}_\perp \right) q_\perp^2 b_p \psi_q \delta_{pq}^k \dd\bs{p} \dd\bs{q}, \\
\label{eq:ikawBzFourier}
k_\perp^2 \pdt b_k - i \Omega_e k_\| k_\perp^2 \psi_k &=& d_e \int_{\mathbb{R}^6} \hat{\bs{e}}_\| \cdot \left(\bs{p}_\perp \times \bs{q}_\perp \right) q_\perp^2 b_p b_q \delta_{pq}^k \dd\bs{p} \dd\bs{q},
\end{eqnarray}
with $\delta_{pq}^k \equiv \delta \left( \bs{k} - \bs{p} - \bs{q}\right)$ the Dirac distribution coming from the Fourier transform of the nonlinear terms. We introduce the canonical variables as follow
\eq{canonicalVariables}{
\psi_k \equiv - \frac{1}{2 d_e k_\perp^2} \sum_{s_k} s_k A_{\bs{k}}^{s_k} \:, \qquad b_k \equiv \frac{1}{2 d_e k_\perp} \sum_{s_k} A_{\bs{k}}^{s_k},
}
where $s_k =\pm$ is the directional polarization that defines the direction of the wave propagation with $s_k k_\| \ge 0$. 
After a little calculation, we find
\eq{}{
\left( \pdt + i s_k \omega_k \right) A_{\bs{k}}^{s_k} = \frac{1}{4} \sum_{s_p s_q} \int_{\mathbb{R}^6} \frac{\hat{\bs{e}}_\| \cdot \left(\bs{p}_\perp \times \bs{q}_\perp\right) }{k_\perp p_\perp q_\perp} \left( q_\perp^2 + s_k s_q k_\perp q_\perp \right) A_{\bs{p}}^{s_p} A_{\bs{q}}^{s_q} \delta_{pq}^k \dd\bs{p} \dd\bs{q}.
}
By making the following change of variable $A_{\bs{k}}^{s_k} = \epsilon a_{\bs{k}}^{s_k} e^{-i s_k \omega_k t}$, where $\epsilon \ll 1$ is a small positive parameter, the linear part of this equation vanishes and we obtain the fundamental equation describing the slow temporal evolution -- thanks to $\epsilon$ -- of the wave amplitude
\eq{kineticFundamental0}{
\pdt  a_{\bs{k}}^{s_k} = \frac{\epsilon}{4}\sum_{s_p s_q} \int_{\mathbb{R}^6} \mathcal{H}_{\bs{kpq}}^{s_k s_p s_q} a_{\bs{p}}^{s_p} a_{\bs{q}}^{s_q} \mathrm{e}^{i \Omega^k_{pq} t} \delta_{pq}^k \dd\bs{p} \dd\bs{q},
}
with $\Omega^k_{pq} \equiv s_k \omega_k - s_p \omega_p - s_q \omega_q$ and $\mathcal{H}_{\bs{kpq}}^{s_k s_p s_q} \equiv \hat{\bs{e}}_\| \cdot \left(\bs{p}_\perp \times \bs{q}_\perp\right) \left( q_\perp^2 + s_k s_q k_\perp q_\perp \right) / \left(k_\perp p_\perp q_\perp \right)$ the nonlinear interaction coefficient which depends on the nonlinearities of the system. 
The presence of the complex exponential is fundamental for the asymptotic closure: as we are interested in the long time behavior with respect to the linear time scale ($1/\omega$), the contribution of the exponential is mostly zero. Only (secular) terms for which $\Omega^k_{pq}=0$ will survive \citep{Benney1966,Newell_2001}. 
Adding to this the relation imposed by the Dirac distribution, we can obtain the following resonance condition (symmetries in $\bs{p}$ and $\bs{q}$ are used)
\begin{eqnarray}
    \bs{k} + \bs{p} + \bs{q} &= \bs{0}, \\
    s_k \omega_k + s_p \omega_p + s_q \omega_q &= 0.
\end{eqnarray}
After a few manipulations, we find the (anisotropic) relationships
\eq{cdtResonnance2}{
\frac{ s_q q_\perp - s_p p_\perp }{s_k \omega_k} = \frac{s_k k_\perp - s_q q_\perp}{s_p \omega_p} = \frac{s_p p_\perp - s_k k_\perp }{s_q \omega_q},
}
which will be useful to prove the conservation of the quadratic invariants. This is also useful to highlight the anisotropic character of the system. Indeed, let us consider the particular case of super-local interactions which give, in general, a dominant contribution to the turbulent dynamics.
In this case, we have $k_\perp \simeq p_\perp \simeq q_\perp$ and the resonance condition simplifies into
\eq{}{
\frac{s_q - s_p}{s_k k_\|} \simeq \frac{s_k - s_q}{s_p p_\|} \simeq \frac{s_p - s_k}{s_q q_\|}.
}
If $k_\|$ is non-zero, the left-hand term will only give a non-negligible contribution when $s_p=-s_q$.
We do not consider the case $s_p=s_q$ which is not relevant to first order in the case of local interactions as can be seen in expression (\ref{eq:kineticFundamental0}) which then becomes negligible (it is easier to see that in equations (\ref{eq:kineticFundamental})--(\ref{eq:coeffInteraction}) after using the symmetry in $\bs{p}$ and $\bs{q}$).
The immediate consequence is that either the middle  or the right term has its numerator canceling (to first order), which implies that the associated denominator must also cancel (to first order) to satisfy the equality: for example, if $s_k=s_p$ then $q_\| \simeq 0$.
This condition means that the transfer in the parallel direction is negligible because the integration in the parallel direction of equation (\ref{eq:kineticFundamental0}) is then reduced to a few modes (since $p_\| \simeq k_\|$) which strongly limits the transfer between the parallel modes.
The cascade in the parallel direction is thus possible but relatively weak compared to the one in the perpendicular direction.

Before applying the spectral formalism of wave turbulence, it is necessary to symmetrize the fundamental equation (\ref{eq:kineticFundamental0})  under the exchange of $\bs{p}$ and $\bs{q}$. 
To to this, we take advantage of the summation over the $s_p$ and $s_q$ polarizations and introduce
\eq{}{
L_{\bs{kpq}}^{s_k s_p s_q} = \frac{1}{2} \left( \mathcal{H}_{\bs{kpq}}^{s_k s_p s_q} + \mathcal{H}_{\bs{kqp}}^{s_k s_q s_p} \right),
}
to finally obtain after a little calculation
\eq{kineticFundamental}{
\pdt a_{\bs{k}}^{s_k} = \epsilon \sum_{s_p s_q} \int_{\mathbb{R}^6} L_{\bs{kpq}}^{s_k s_p s_q} a_{\bs{p}}^{s_p} a_{\bs{q}}^{s_q} \mathrm{e}^{i \Omega_{pq}^k t} \delta_{pq}^k \dd\bs{p} \dd\bs{q},
}
where
\eq{coeffInteraction}{
L_{\bs{kpq}}^{s_k s_p s_q} \equiv \frac{
\hat{\bs{e}}_\| \cdot \left(\bs{p}_\perp \times \bs{q}_\perp
\right) }
{8k_\perp p_\perp q_\perp} \left(s_q q_\perp - s_p p_\perp \right) \left( s_k k_\perp +  s_p p_\perp + s_q q_\perp \right).
}
This operator has, among others, the following symmetries
\begin{eqnarray}
L_{\bs{kpq}}^{s_k s_p s_q} &=& L_{\bs{kpq}}^{-s_k -s_p -s_q} = L_{\bs{-k-p-q}}^{s_k s_p s_q} = L_{\bs{-kpq}}^{s_k s_p s_q} =  L_{\bs{k-p-q}}^{s_k s_p s_q}, \\
L_{\bs{kpq}}^{s_k s_p s_q} &=& L_{\bs{kqp}}^{s_k s_q s_p}, \\
L_{\bs{kpq}}^{s_k -s_p -s_q}  &=& L_{\bs{kpq}}^{-s_k s_p s_q},\\
L_{\bs{0pq}}^{s_k s_p s_q} &=& 0.
\end{eqnarray}
Equation (\ref{eq:kineticFundamental}) is our fundamental equation, the starting point to derive the wave kinetic equations. Note that the nonlinear coupling associated with the wavevectors $\bs{p}$ and $\bs{q}$ vanishes when they are collinear ($k=0$ is a particular case).  
Additionally, the nonlinear coupling vanishes whenever the wavenumbers $p_\perp$ and $q_\perp$ are equal if their associated polarities $s_p$ and $s_q$ are also equal. This was also observed in EMHD (for scales larger than $d_e$) and seems to be a general property of helical waves \citep{Kraichnan1973,Waleffe1992,Turner2000,Galtier_2003a,Galtier_2003b}.

\section{Phenomenology of wave turbulence}\label{sec:Phenomenology}

Before going into the deep analysis of the wave turbulence regime, it is important to have a simple (phenomenological) picture in mind of the physical process that we are going to describe.
According to the properties given in section \ref{sec:Triadic}, if we assume that the nonlinear transfer is mainly driven by super-local interactions $\left( k \sim p \sim q \right)$, which is a classical assumption in the turbulence phenomenology, then we can consider only stochastic collisions between counter propagating waves $(s_p = - s_q)$ to derive the form of the spectra. 
Note that non-local interactions (which include copropagating waves) also provide a contribution to the nonlinear dynamics but, as will be shown in section 6.4 with the convergence study, their contributions are not dominant for the formation of a stationary spectrum.

To find the transfer time and then the energy spectrum, we first need to evaluate the modification of a wave produced by one collision. Starting from the momentum equation (for simplicity we write the wave amplitude as $a_\ell$ and assume anisotropy with $k \sim k_\perp$), we have
\eq{}{
a_\ell(t+\tau_1) \sim a_\ell(t) + \tau_1 \frac{\partial a}{\partial t} \sim a_\ell(t) + \tau_1 \frac{a_\ell^2}{\ell_\perp} ,
}
where $\tau_1$ is the duration of one collision; in other words, after a collision, the distortion of a wave is $\Delta_1 a_\ell \simeq a_\ell^2 / \ell_\perp$. This distortion is going to increase with time in such a way that after $N$ stochastic collisions, the cumulative effect may be evaluated like a random walk \citep{Galtier_2016}
\eq{}{
\sum_{i=1}^N \Delta_i a_\ell \sim \tau_1 \frac{a_\ell^2}{\ell_\perp} \sqrt{\frac{t}{\tau_1}}.
}
The transfer time, $\tau_\mathrm{tr}$, that we are looking for is the time for which the cumulative distortion is of order one, i.e. of the order of the wave itself:
\eq{}{
a_\ell \sim \tau_1 \frac{a_\ell^2}{\ell_\perp} \sqrt{\frac{\tau_\mathrm{tr}}{\tau_1}}.
}
Then, we obtain
\eq{}{
\tau_\mathrm{tr} \sim \frac{1}{\tau_1} \frac{\ell_\perp^2}{a_\ell^2} \sim \frac{\tau_\mathrm{NL}^2}{\tau_1},
}
where $\tau_\mathrm{NL} \equiv \ell_\perp / a_\ell$. This is basically the formula that we are going to use to evaluate the energy spectra. Let us consider IWW/IKAW for which $\tau_1 \sim 1/\omega_k \sim k_\perp/ k_\parallel$. A classical calculation with a constant energy flux $\varepsilon \sim E_\ell / \tau_\mathrm{tr}$, leads finally to the bi-dimensionnal axisymmetric energy spectrum
\eq{}{
E \left( k_\perp, k_\| \right) \sim \sqrt{\varepsilon \Omega_e} k_\perp^{-5/2} k_\|^{-1/2}.
}
As we will see in § \ref{subsec:KZKspectra}, this corresponds to the exact solution of the wave turbulence theory. 
The same calculation could be done for the momentum but, as we will see, it presents a more subtle behavior that the phenomenology cannot describe.

\section{Kinetic equations}\label{sec:Kinetic}

\subsection{Definition of the energy density tensor}

We now move on to a statistical description. We use the ensemble average $\langle \rangle$ and define the following spectral correlators (cumulants) for homogeneous turbulence (we assume $\langle a_k^{s_k}\rangle=0$)
\eq{}{
\left\langle a_{\bs{k}}^{s_k} a_{\bs{k'}}^{s_k'} \right\rangle = e_{\bs{k}'}^{s_k'} \delta_{kk'}\delta^{s_k}_{s_k'},
}
with $e^{s_k'}\left( \bs{k}'\right) = e_{\bs{k}'}^{s_k'}$. 
We observe the presence of the delta function $\delta^{s_k}_{s_k'}$  meaning that two-point correlations of opposite polarities have no long-time influence in the wave turbulence regime. The other delta function is the consequence of the statistical homogeneity assumption. 
The objective of the wave turbulence theory is to derive a  self-consistent equation for the time evolution of this spectral correlator; this is the kinetic equation. In this development, we have to face the classical closure problem: a hierarchy of statistical equations of increasingly higher order emerges. In contrast to strong turbulence, in the weak wave turbulence regime we can use the time scale separation to achieve a natural closure of the system \citep{Benney1966,Newell_2001}. 
After a lengthy (but classical) algebra, we obtain the time evolution equation of the energy density tensor (we leave the details of the derivation to Appendix \ref{appendix:Moments}) 
\eq{densityEnergyEvolution}{
\begin{split}
\pdt e_{\bs{k}}^{s_k} = \frac{\pi \epsilon^2}{16} \sum_{s_p s_q} \int_{\mathbb{R}^6} & s_k \omega_k \left\vert \tilde L_{\bs{kpq}}^{s_k s_p s_q} \right\vert^2 \left( s_k \omega_k e_{\bs{p}}^{s_p} e_{\bs{q}}^{s_q} + s_p \omega_p e_{\bs{k}}^{s_k} e_{\bs{q}}^{s_q} + s_q \omega_q e_{\bs{k}}^{s_k} e_{\bs{p}}^{s_p}
\right) \\ 
&\times \delta\left(\Omega_{kpq} \right) \delta_{kpq} \dd\bs{p} \dd\bs{q},
\end{split}
}
where 
\eq{Ltilde}{
\tilde L_{\bs{kpq}}^{s_k s_p s_q} \equiv
 \frac{L_{\bs{kpq}}^{s_k s_p s_q}}{s_k \omega_k} ,
}
$\Omega_{kpq} \equiv s_k \omega_k + s_p \omega_p + s_q \omega_q$ and $\delta_{kpq} =  \delta \left(\bs{k} + \bs{p} + \bs{q}\right)$. This equation is the main result of the wave turbulence formalism. It describes the statistical properties of IWW or IKAW turbulence at the leading order, i.e. for three-wave interactions.

\subsection{Detailed conservation of quadratic invariants}

In § \ref{subsec:invariants} we introduced the three-dimensional invariants of IEMHD. The first test that the wave turbulence equations must pass is the detailed conservation -- i.e. for each triad ($\bs{k}, \bs{p}, \bs{q}$) -- of these invariants. Starting from the definitions (\ref{eq:totalEnergy}) and (\ref{eq:totalHelicity}), we define the  energy and momentum spectra
\begin{eqnarray}
    E(\bs{k}) &\equiv& \sum_{s_k=\pm} e_{\bs{k}}^{s_k} = e_{\bs{k}}^+ + e_{\bs{k}}^-, \\
    H(\bs{k}) &\equiv& \sum_{s_k=\pm} s_k k_\perp e_{\bs{k}}^{s_k} = k_\perp \left( e_{\bs{k}}^+ - e_{\bs{k}}^- \right).
\end{eqnarray}
Before checking the energy conservation, it is interesting to note that when one of the polarized energy density tensors $e_{\bs{k}}^\pm$ is zero, the other invariant is extremal and verifies the relation $ H(\bs{k}) = \pm k_\perp E(\bs{k})$, which is in agreement with the realizability condition
(Schwarz inequality) 
$ \vert H(\bs{k}) \vert \le k_\perp E(\bs{k})$.
From equation (\ref{eq:densityEnergyEvolution}), we obtain the equation for the (total) energy
\eq{densityEnergy}{
\begin{split}
     \pdt E(t) &\equiv \pdt \int_{\mathbb{R}^3} \sum_{s_k} e_{\bs{k}}^{s_k} \dd\bs{k} \\
&= \frac{\pi \epsilon^2}{16}  \sum_{s_k s_p s_q} \int_{\mathbb{R}^9} 
s_k \omega_k
\left\vert \tilde L_{\bs{kpq}}^{s_k s_p s_q} \right\vert^2 \left( s_k \omega_k e_{\bs{p}}^{s_p} e_{\bs{q}}^{s_q} + s_p \omega_p e_{\bs{k}}^{s_k} e_{\bs{q}}^{s_q} + s_q \omega_q e_{\bs{k}}^{s_k} e_{\bs{p}}^{s_p} \right) \\
&\hspace{3cm} \times \delta\left(\Omega_{kpq} \right) \delta_{kpq} \dd\bs{k} \dd\bs{p} \dd\bs{q}.
\end{split}
}
Without forcing and dissipation, energy must be conserved and this conservation is done at the level of triadic interactions (detailed energy conservation). The demonstration is straightforward. By applying a cyclic permutation of wavevectors and polarizations, we find
\eq{}{
\begin{split}
\pdt E(t) = \frac{\pi \epsilon^2}{48}  \sum_{s_k s_p s_q} \int_{\mathbb{R}^9} &
\Omega_{kpq} s_k \omega_k
\left\vert \tilde L_{\bs{kpq}}^{s_k s_p s_q} \right\vert^2 \left( s_k \omega_k e_{\bs{p}}^{s_p} e_{\bs{q}}^{s_q} + s_p \omega_p e_{\bs{k}}^{s_k} e_{\bs{q}}^{s_q} + s_q \omega_q e_{\bs{k}}^{s_k} e_{\bs{p}}^{s_p} \right) \\
&\times \delta\left(\Omega_{kpq} \right) \delta_{kpq} \dd\bs{k} \dd\bs{p} \dd\bs{q} = 0 , 
\end{split}
}
which proves the conservation of (kinetic) energy on the resonant manifold for each triadic interaction. 

For the second invariant $H(t)$, one has
\eq{densityHelicity}{
\begin{split}
\pdt H(t) &\equiv \pdt \int_{\mathbb{R}^3} \sum_{s_k} s_k k_\perp e_{\bs{k}}^{s_k} \dd\bs{k} \\
&= \frac{\pi \epsilon^2 \Omega_e}{16}  \sum_{s_k s_p s_q} \int_{\mathbb{R}^9}  k_\| \left\vert \tilde L_{\bs{kpq}}^{s_k s_p s_q} \right\vert^2 \left( s_k \omega_k e_{\bs{p}}^{s_p} e_{\bs{q}}^{s_q} + s_p \omega_p e_{\bs{k}}^{s_k} e_{\bs{q}}^{s_q} + s_q \omega_q e_{\bs{k}}^{s_k} e_{\bs{p}}^{s_p} \right) \\
&\hspace{3cm} \times \delta\left(\Omega_{kpq} \right) \delta_{kpq} \dd\bs{k} \dd\bs{p} \dd\bs{q}.
\end{split}
}
The same manipulations as before leads immediately to
\eq{}{
\begin{split}
\pdt H(t) = \frac{\pi \epsilon^2 \Omega_e}{48}  \sum_{s_k s_p s_q} \int_{\mathbb{R}^9} &\left\vert \tilde L_{\bs{kpq}}^{s_k s_p s_q} \right\vert^2 \left( s_k \omega_k e_{\bs{p}}^{s_p} e_{\bs{q}}^{s_q} + s_p \omega_p e_{\bs{k}}^{s_k} e_{\bs{q}}^{s_q} + s_q \omega_q e_{\bs{k}}^{s_k} e_{\bs{p}}^{s_p} \right) \\
&\times \left( k_\| + p_\| + q_\| \right)  \delta\left(\Omega_{kpq} \right) \delta_{kpq} \dd\bs{k} \dd\bs{p} \dd\bs{q} = 0.
\end{split}
}
This proves the conservation of momentum (kinetic helicity) on the resonant manifold for each triadic interaction.

\subsection{Helical turbulence}
From the wave turbulence equation (\ref{eq:densityEnergyEvolution}), we can deduce several general properties. 
First, we observe that there is no coupling between the waves associated with the $\bs{p}$ and $\bs{q}$ wavevectors when these wavevectors are collinear. Second, the nonlinear coupling disappears whenever the wavenumbers $p_\perp$ and $q_\perp$ are equal if their associated polarities $s_p$ and $s_q$ are also equal. These properties are also observed in EMHD (for scales larger than $d_e$) and more generally for other helical waves \citep{Kraichnan1973,Waleffe1992,Turner2000,Galtier_2003a,Galtier_2003b}.
Note that they can already be deduced directly from the fundamental equation (\ref{eq:kineticFundamental0}).
Third, the wave modes ($k_\| > 0$) are decoupled from the slow mode ($k_\|=0$) which is not described by these wave kinetic equations. This situation is thus different from wave turbulence in incompressible MHD where the slow mode has a profound influence on the nonlinear dynamics.

\section{Turbulent spectra as exact solutions}\label{sec:Spectra}

\subsection{Wave kinetic equations for the invariants}

The objective of this section is to derive, in the stationary case, the exact power law solutions of the kinetic equations for the two invariants, energy and momentum. 
To do so, it is necessary to simplify the equations, written for $E(\bs{k})$ and $H(\bs{k})$, using the axisymmetric assumption. First of all, we have
\eq{kineticInvariants}{
\begin{split}
\pdt 
\begin{pmatrix}
E(\bs{k}) \\
H(\bs{k}) 
\end{pmatrix}
 = \frac{\pi \epsilon^2}{16}  \sum_{s_k s_p s_q} \int_{\mathbb{R}^6} &s_k \omega_k \left\vert \tilde L_{\bs{kpq}}^{s_k s_p s_q} \right\vert^2
\left( s_k \omega_k e_{\bs{p}}^{s_p} e_{\bs{q}}^{s_q} + s_p \omega_p e_{\bs{k}}^{s_k} e_{\bs{q}}^{s_q} + s_q \omega_q e_{\bs{k}}^{s_k} e_{\bs{p}}^{s_p} \right) \\
& \times \begin{pmatrix}
1 \\
s_k k_\perp
\end{pmatrix}
\delta\left(\Omega_{kpq} \right) \delta_{kpq} \dd\bs{p} \dd\bs{q}.
\end{split}
}
We now develop the energy density tensors inside the integral in terms of energy and momentum spectra. We note that only terms containing the products of two $E(\bs{k})$ or two $H(\bs{k})$ will survive for energy, whereas only the products of $E(\bs{k}) H(\bs{k})$ will survive for helicity. After some algebra, we find for the energy
\eq{}{
\begin{split}
\pdt 
E(\bs{k}) =& \frac{\pi \epsilon^2}{64} \sum_{s_k s_p s_q} \int_{\mathbb{R}^6} s_k \omega_k \left\vert \tilde L_{\bs{kpq}}^{s_k s_p s_q} \right\vert^2 \\
&\times \left[ s_k \omega_k E (\bs{p}) E (\bs{q}) + s_p \omega_p E (\bs{k}) E (\bs{q}) + s_q \omega_q E (\bs{k}) E (\bs{p}) \right. \\
& \left. + s_k s_p s_q \left( \omega_k\frac{H(\bs{p}) H(\bs{q})}{p_\perp q_\perp} + \omega_p \frac{H(\bs{k}) H(\bs{q})}{k_\perp q_\perp} + \omega_q\frac{H(\bs{k}) H(\bs{p})}{k_\perp p_\perp} \right) \right] \\
&\times \delta \left(\Omega_{kpq} \right) \delta_{kpq} \dd\bs{p} \dd\bs{q},
\end{split}
}
and for the momentum
\eq{}{
\begin{split}
\pdt 
H(\bs{k})
 = \frac{\pi \epsilon^2}{64} \sum_{s_k s_p s_q} \int_{\mathbb{R}^6} &\omega_k k_\perp \left\vert \tilde L_{\bs{kpq}}^{s_k s_p s_q} \right\vert^2 \left(
s_k \omega_k \left[ E(\bs{p}) \frac{H(\bs{q})}{s_q q_\perp} + E(\bs{q}) \frac{H(\bs{p})}{s_p p_\perp} \right] \right.\\
&\left. +
s_p \omega_p \left[ E(\bs{k}) \frac{H(\bs{q})}{s_q q_\perp} + E(\bs{q}) \frac{H(\bs{k})}{s_k k_\perp} \right] \right. \\
&\left. +
s_q \omega_q \left[E(\bs{k}) \frac{H(\bs{p})}{s_p p_\perp} + E(\bs{p}) \frac{H(\bs{k})}{s_k k_\perp} \right]
\right)
\delta\left(\Omega_{kpq} \right) \delta_{kpq} \dd\bs{p} \dd\bs{q}.
\end{split}
}
If we exchange in the integrand the dummy variables, $\bs{p}$ and $\bs{q}$, as well as $s_p$ and $s_q$ , we can simplify further the previous expressions to obtain
\eq{kineticEH}{
\pdt 
\left( E(\bs{k}) \atop H(\bs{k}) \right)
 = \frac{\pi \epsilon^2}{32} \sum_{s_k s_p s_q} \int_{\mathbb{R}^6} \left\vert \tilde L_{\bs{kpq}}^{s_k s_p s_q} \right\vert^2
s_k \omega_k s_p \omega_p
\left( X_E \atop X_H \right)
\delta\left(\Omega_{kpq} \right) \delta_{kpq} \dd\bs{p} \dd\bs{q},
}
with
\eq{}{
\left( X_E \atop  X_H \right)
=
\left(
E (\bs{q}) \left[ E (\bs{k}) - E (\bs{p}) \right]
+
\frac{H(\bs{q})}{s_q q_\perp} \left( \frac{H(\bs{k})}{s_k k_\perp} - \frac{H(\bs{p})}{s_p p_\perp} \right) \atop
s_k k_\perp \left[ E(\bs{q}) \left( \frac{H(\bs{k})}{s_k k_\perp} - \frac{H(\bs{p})}{s_p p_\perp} \right)
+
\frac{H(\bs{q})}{s_q q_\perp} \left[ E(\bs{k}) - E(\bs{p})\right] \right]
\right).
}

\subsection{The axisymmetric wave turbulence equations}
 
\begin{figure}
    \centering
\tikzset{every picture/.style={line width=0.75pt}} 
\begin{tikzpicture}[x=0.75pt,y=0.75pt,yscale=-1,xscale=1]
\draw    (321.08,26.36) -- (272.74,156.06) ;
\draw [shift={(271.69,158.87)}, rotate = 290.44] [fill={rgb, 255:red, 0; green, 0; blue, 0 }  ][line width=0.08]  [draw opacity=0] (8.93,-4.29) -- (0,0) -- (8.93,4.29) -- cycle    ;
\draw    (271.69,158.87) -- (382.9,144.46) ;
\draw [shift={(385.87,144.08)}, rotate = 532.62] [fill={rgb, 255:red, 0; green, 0; blue, 0 }  ][line width=0.08]  [draw opacity=0] (8.93,-4.29) -- (0,0) -- (8.93,4.29) -- cycle    ;
\draw    (385.87,144.08) -- (322.53,28.99) ;
\draw [shift={(321.08,26.36)}, rotate = 421.16999999999996] [fill={rgb, 255:red, 0; green, 0; blue, 0 }  ][line width=0.08]  [draw opacity=0] (8.93,-4.29) -- (0,0) -- (8.93,4.29) -- cycle    ;

\draw  [draw opacity=0] (332.39,46.13) .. controls (329.91,47.63) and (327.05,48.51) .. (323.99,48.56) .. controls (320.55,48.63) and (317.31,47.63) .. (314.57,45.86) -- (323.66,30.07) -- cycle ; \draw   (332.39,46.13) .. controls (329.91,47.63) and (327.05,48.51) .. (323.99,48.56) .. controls (320.55,48.63) and (317.31,47.63) .. (314.57,45.86) ;
\draw  [draw opacity=0] (278.63,141.64) .. controls (281.22,142.94) and (283.49,144.9) .. (285.16,147.46) .. controls (287.04,150.35) and (287.91,153.62) .. (287.87,156.88) -- (269.66,157.57) -- cycle ; \draw   (278.63,141.64) .. controls (281.22,142.94) and (283.49,144.9) .. (285.16,147.46) .. controls (287.04,150.35) and (287.91,153.62) .. (287.87,156.88) ;
\draw  [draw opacity=0] (368.88,146.33) .. controls (368.78,143.43) and (369.41,140.51) .. (370.86,137.81) .. controls (372.49,134.77) and (374.94,132.44) .. (377.82,130.91) -- (387.16,146.57) -- cycle ; \draw   (368.88,146.33) .. controls (368.78,143.43) and (369.41,140.51) .. (370.86,137.81) .. controls (372.49,134.77) and (374.94,132.44) .. (377.82,130.91) ;

\draw (330,157.4) node [anchor=north west][inner sep=0.75pt] {$\boldsymbol{k}_\perp$};
\draw (365,80) node [anchor=north west][inner sep=0.75pt] {$\boldsymbol{p}_\perp$};
\draw (280,80) node [anchor=north west][inner sep=0.75pt] {$\boldsymbol{q}_\perp$};
\draw (314.08,54.76) node [anchor=north west][inner sep=0.75pt] {$\alpha_{k}$};
\draw (288.08,129.76) node [anchor=north west][inner sep=0.75pt] {$\alpha_{p}$};
\draw (345.08,117.76) node [anchor=north west][inner sep=0.75pt] {$\alpha_{q}$};

\end{tikzpicture}
    \caption{Triadic relation $\bs{k}_\perp + \bs{p}_\perp + \bs{q}_\perp = \bs{0}$.}
    \label{fig:resonantCondition}
\end{figure}
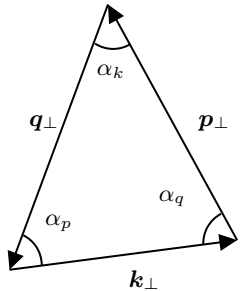

To simplify the problem, we will consider an axial symmetry with respect to the external magnetic field and introduce the two-dimensional anisotropic spectra
\begin{eqnarray}
    E_k &= E \left( k_\perp, k_\| \right) &= 2\pi k_\perp E \left( \bs{k}_\perp, k_\| \right), \\
    H_k &= H \left( k_\perp, k_\| \right) &= 2\pi k_\perp H \left( \bs{k}_\perp, k_\| \right),
\end{eqnarray}
which result from an integration over the angles in the plane perpendicular to the mean magnetic field (see Figure \ref{fig:resonantCondition}). In polar coordinates $\dd\bs{p} \dd\bs{q} = p_\perp \dd \alpha_q \dd p_\perp \dd p_\| \dd q_\|$ and, thanks to the Al-Kashi formula: $q_\perp^2 = k_\perp^2 + p_\perp^2 - 2 k_\perp p_\perp \cos \alpha_q$, we find at fixed $k_\perp$ and $p_\perp$, $q_\perp \dd q_\perp = k_\perp p_\perp \sin \alpha_q \dd \alpha_q$. Using expression (\ref{eq:Ltilde}), we then obtain the kinetic equations
\eq{EHcinetiqueLocal}{
\begin{split}
\pdt 
\begin{pmatrix}
E_k \\
H_k 
\end{pmatrix}
 = \frac{\epsilon^2 \Omega_e^2}{2^{12}}  \sum_{s_k s_p s_q} \int_{\Delta_\perp} &\frac{s_k s_p k_\| p_\|}{k_\perp^2 p_\perp^2 q_\perp^2} \left(\frac{s_q q_\perp - s_p p_\perp}{ s_k \omega_k}\right)^2 \left( s_k k_\perp + s_p p_\perp + s_q q_\perp \right)^2 \sin \alpha_q \\
&\times 
\begin{pmatrix}
\tilde{X}_E \\
\tilde{X}_H
\end{pmatrix}
\delta \left(\Omega_{kpq} \right) \delta_{k_\|p_\|q_\|} \dd p_\perp \dd q_\perp \dd p_\| \dd q_\|,
\end{split}
}
where $\Delta_\perp$ the integration domain verifies the resonance condition $\bs{k}_\perp + \bs{p}_\perp + \bs{q}_\perp = \bs{0} $ and
\eq{toto}{
\begin{pmatrix}
\tilde{X}_E \\
\tilde{X}_H
\end{pmatrix}
=
\begin{pmatrix}
E_q \left( p_\perp E_k - k_\perp E_p \right)
+
\frac{H_q}{s_q q_\perp} \left( \frac{p_\perp}{s_k k_\perp}H_k - \frac{k_\perp}{s_p p_\perp}H_p \right) \\
s_k k_\perp \left[ E_q \left( \frac{p_\perp}{s_k k_\perp}H_k - \frac{k_\perp}{s_p p_\perp}H_p \right)
+
\frac{H_q}{s_q q_\perp}  \left( p_\perp E_k - k_\perp E_p \right) \right] 
\end{pmatrix} ,
}
with $\alpha_q$ the angle between $\bs{k}_\perp$ and $\bs{p}_\perp$ in the triangle defined by the triadic  interaction $( \bs{k}_\perp, \bs{p}_\perp, \bs{q}_\perp)$ (see Figure \ref{fig:resonantCondition}).
Equations (\ref{eq:EHcinetiqueLocal}) will be 
used to derive exact solutions also called  Kolmogorov-Zakharov spectra. 

\subsection{Kolmogorov-Zakharov spectra} \label{subsec:KZKspectra}

Equations (\ref{eq:EHcinetiqueLocal}) have sufficient symmetry to apply the bi-homogeneous conformal Kuznetsov–Zakharov transformation \citep{Zakharov_1992}. This transformation has been applied to several problems involving anisotropy \citep{Kuznetsov_2001,Galtier_2003a,Galtier_2006}. It is a generalization of the Zakharov transformation applied for isotropic turbulence (in the context of strong 2D HD turbulence, see also \cite{Kraichnan_1967}). 
With such an operation, we are able to find the exact stationary solutions of the kinetic equations in power law form.
The bihomogeneity of the integrals in the wavenumbers $k_\perp$ and $k_\|$ allows us to use the transformations (see Figure \ref{fig:KZKTransformation})
\begin{eqnarray}
p_\perp &\to& k_\perp^2/ p_\perp,\\
q_\perp &\to& k_\perp q_\perp/ p_\perp,\\
p_\| &\to& k_\|^2/ p_\|,\\
q_\| &\to& k_\| q_\| / p_\|. 
\end{eqnarray}
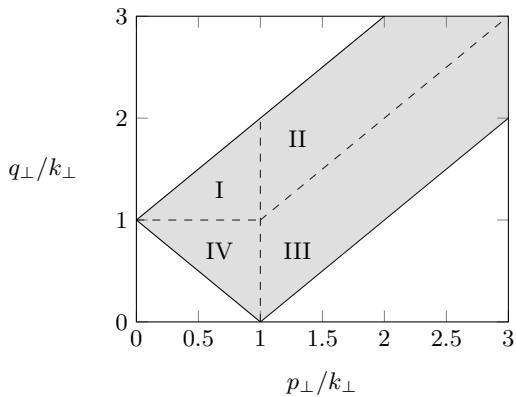
\begin{figure}
    \centering
    \begin{tikzpicture}
        \begin{axis}[small,
            ymin=0,ymax=3,xmin=0,xmax=3,
            xlabel=$p_\perp / k_\perp$, ylabel=$q_\perp / k_\perp$, ylabel style={rotate=-90}
            ]
            \addplot[domain=0:5]{x-1};
            \addplot[domain=1:5, dashed]{x};
            \addplot[domain=0:1, dashed]{1};
            \addplot[domain=0:1, dashed] coordinates {(1,0)(1,2)};
            \addplot [fill=gray, fill opacity=0.25] coordinates {(1, 0)(0, 1)(4, 5)(5, 4)};
            \node[] at (axis cs: 0.67,1.3) {I};
            \node[] at (axis cs: 1.3,1.8) {II};
            \node[] at (axis cs: 1.3,0.7) {III};
            \node[] at (axis cs: 0.67,0.7) {IV};
        \end{axis}
    \end{tikzpicture}
    \caption{Illustration of the Kuznetsov-Zakharov transformation. It consists in swapping regions I and III with regions II and IV, respectively. We specify that the gray band is defined up to infinity and corresponds to the domain $\Delta_\perp$. The same manipulation is done on the parallel wavenumbers.}
    \label{fig:KZKTransformation}
\end{figure}
We apply this transformation first on the energy equation (\ref{eq:EHcinetiqueLocal}) which means that we are looking for constant energy flux solutions. 
We seek stationary solutions in the power law form, 
\eq{plsol}{
E \left(k_\perp, k_\| \right) = C_E k_\perp^{-x} \left\vert k_\| \right\vert ^{-y} 
\quad {\rm and} \quad 
H \left( k_\perp, k_\| \right) = C_H k_\perp^{-\tilde{x}} \left\vert k_\| \right\vert^{-\tilde{y}} ,
}
where $C_E$ and $C_H$ are two constants with $C_E \ge 0$. (We consider only positive parallel wavenumber since it is symmetric in $k_\|$.) 
The new form of the integral, resulting from the summation of the integrand in its primary form and after the Kuznetsov–Zakharov transformation, can be written as
\eq{energyKZKsolution}{
\begin{split}
    \pdt E_k = \frac{\epsilon^2 \Omega_e^2}{2^{13}} \sum_{s_k s_p s_q} \int_{\Delta_\perp} &\frac{s_k s_p k_\| p_\|}{k_\perp^2 p_\perp q_\perp^2} \left(\frac{s_q q_\perp - s_p p_\perp}{ s_k \omega_k}\right)^2 \left( s_k k_\perp + s_p p_\perp + s_q q_\perp \right)^2 \sin \alpha_q \\
    &\times \left(  C_E^2  \Xi_E + s_k s_q C_H^2 \Xi_H\right) \delta \left(\Omega_{kpq} \right) \delta_{k_\|p_\|q_\|}  \dd p_\perp \dd q_\perp \dd p_\| \dd q_\|,
\end{split}
}
with the pure energy contribution
\eq{}{
\Xi_E = k_\perp^{-x} {\left\vert k_\| \right\vert}^{-y} q_\perp^{-x} {\left\vert q_\| \right\vert}^{-y} \left[ 1 - \left( \frac{p_\perp}{k_\perp}\right)^{-1-x} \left\vert \frac{p_\|}{k_\|}\right\vert^{-y} \right]\left[ 1 - \left( \frac{p_\perp}{k_\perp}\right)^{-5+2x} \left\vert \frac{p_\|}{k_\|}\right\vert^{-1+2y} \right],
}
and the pure helicity contribution
\eq{}{
    \Xi_H = k_\perp^{-1-\tilde{x}} {\left\vert k_\| \right\vert}^{-\tilde{y}} q_\perp^{-1-\tilde{x}} {\left\vert q_\| \right\vert}^{-\tilde{y}} \left[ 1  - s_k s_p \left( \frac{p_\perp}{k_\perp}\right)^{-\tilde{x}-2} \left\vert \frac{p_\|}{k_\|}\right\vert^{-\tilde{y}}\right] \left[ 1 - \left( \frac{p_\perp}{k_\perp}\right)^{2\tilde{x}-3} \left\vert \frac{p_\|}{k_\|}\right\vert^{2\tilde{y}-1} \right].
}
We can distinguish two different types of solutions. First, there are the thermodynamic equilibrium solutions, which correspond to the equipartition state for which the energy flux is zero.
The power laws which verify this condition are 
\begin{eqnarray}
    E \left(k_\perp, k_\| \right) &=& C_E k_\perp, \\
    H \left(k_\perp, k_\| \right) &=& C_H k_\perp^2.
\end{eqnarray}
These results can be easily verified by a direct substitution in the original kinetic equations.
In general, this stationary state cannot be reached in the presence of helicity because the value $s_k s_p = -1$ prevents the cancellation of the integral. 
There is, however, a particular case where the solutions exist: it is the state of maximal helicity for which either  $e_{\bs{k}}^+=0$ or $e_{\bs{k}}^-=0$. Then, we have the relation $H_k=\pm k_\perp E_k$. But this state is not viable as we can see on equation (\ref{eq:densityEnergyEvolution}): for example, if $e_{\bs{k}}^-=0$ at time $t=0$, it will not remain zero at time $t>0$. This means that this solution is only possible if there is an external mechanism that forces the system to remain in the maximal helicity state. 

The most interesting solutions are those for which the energy flux is constant, non-zero and finite.
These exact solutions are called Kolmogorov-Zakharov (KZ) spectra and correspond to the values which make the integral cancels in a non-trivial way and independently of the polarizations. These spectra are
\begin{eqnarray}
    \label{eq:energySolutionKZ}
    E \left(k_\perp, k_\| \right) &=& C_E k_\perp^{-5/2} { \left\vert k_\| \right\vert^{-1/2}}, \\
     \label{eq:helicitySolutionKZ}
    H \left(k_\perp, k_\| \right) &=& C_H k_\perp^{-3/2} { \left\vert k_\| \right\vert^{-1/2}}.
\end{eqnarray}
There are not constrained by the polarization and can  therefore be reached by the system even in the presence of helicity. 

For the helicity equation, using the same manipulations as before, we obtain
\eq{helicitySpectra}{
\begin{split}
    \pdt H_k =& \frac{\epsilon^2 \Omega_e^2}{{2^{13}}} C_E C_H \sum_{s_k s_p s_q} \int_{\Delta_\perp} \frac{s_k s_p k_\| p_\| }{k_\perp^2 p_\perp q_\perp^2} \left(\frac{s_q q_\perp - s_p p_\perp}{ s_k \omega_k}\right)^2 \left( s_k k_\perp + s_p p_\perp + s_q q_\perp \right)^2 \sin \alpha_q \\
    &\times k_\perp^{-x-\tilde{x}} \left\vert k_\| \right\vert^{-y-\tilde{y}} \left[1 - s_k s_p \left(\frac{p_\perp}{k_\perp}\right)^{x+\tilde{x}-4} \left\vert \frac{p_\|}{k_\|}\right\vert^{y+\tilde{y}-1} \right] \\
    &\times \left[\left(\frac{q_\perp}{k_\perp}\right)^{-x} \left\vert \frac{q_\|}{k_\|}\right\vert^{-y} \left(1 - s_k s_p \left(\frac{p_\perp}{k_\perp}\right)^{-\tilde{x}-2} \left\vert \frac{p_\|}{k_\|}\right\vert^{-\tilde{y}} \right) \right. \\
    &\left. + s_k s_q \left(\frac{q_\perp}{k_\perp}\right)^{-\tilde{x}-1} \left\vert \frac{q_\|}{k_\|}\right\vert^{-\tilde{y}} \left( 1 - \left(\frac{p_\perp}{k_\perp}\right)^{-x-1} \left\vert \frac{p_\|}{k_\|}\right\vert^{-y} \right)\right] \delta \left(\Omega_{kpq} \right) \delta_{k_\|p_\|q_\|}  \dd p_\perp \dd q_\perp \dd p_\| \dd q_\|.
\end{split}
}
The zero helicity flux solutions satisfy \begin{eqnarray}
    E \left(k_\perp, k_\| \right) &=& C_E k_\perp, \\
    H \left(k_\perp, k_\| \right) &=& C_H k_\perp^2,
\end{eqnarray}
which correspond to the thermodynamic spectra found for energy (this can be seen directly from equation (\ref{eq:toto})). 
For the KZ spectra, we have a family of solutions that meet the following criteria 
\begin{eqnarray}
    x + \tilde{x} &=& 4, \\
    y + \tilde{y} &=& 1.
\end{eqnarray}
The situation is worse than for energy because none of the constant helicity flux solutions (thermodynamic or KZ) can be reached in general because of the presence of the product $s_ks_p$ which, let us recall, prevents the cancellation of the term in the right-hand side of expression (\ref{eq:helicitySpectra}). Only the maximal helicity state allows the existence of these stationary spectra but, as said above, it is not a naturally viable state. 
(Note that this property found in weak wave turbulence may not be true in strong turbulence.)

In conclusion, the most relevant solutions are the KZ spectra at constant energy flux. In section \ref{sec:Flux} we will further investigate the corresponding exact solution for $H=0$ in order to find the direction of the energy cascade and the expression of the Kolmogorov constant.  
In space plasma physics, we often compare theoretical predictions with the magnetic spectrum $E_k^B$ which is well measured by spacecraft (with the Taylor hypothesis, the frequency is used as a proxy for the wavenumber). In our case, a simple dimensional analysis based on the definition of energy (\ref{eq:totalEnergy}), leads to the relation $E_k \sim k_\perp^2 E_k^B$. Consequently, we obtain $E_k^B \sim k_\perp^{-9/2}$, which is steeper than the predictions made at scales larger than $d_e$.

\subsection{Locality condition}
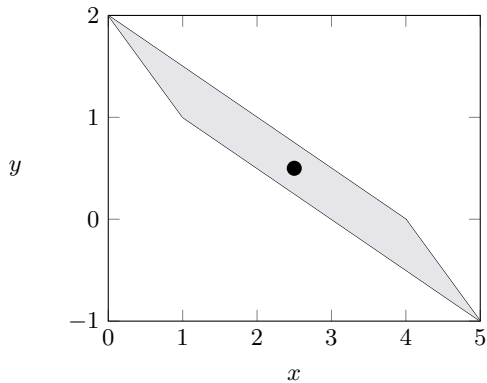
\begin{figure}
    \centering
    \begin{tikzpicture}
        \begin{axis}[small,
            ymin=-1,ymax=2,xmin=0,xmax=5,
            xlabel=$x$, ylabel=$y$, ylabel style={rotate=-90}
            ]
            \addplot[name path global=firstline,  domain=0:4.01]{2-x/2};
            \addplot[name path global=secondline, domain=4:5]{4-x};
            \addplot[name path global=thirdline,  domain=1:5]{3/2-x/2};
            \addplot[name path global=fourthline, domain=0:1.001]{2-x};
            \fill[name intersections={of=firstline and secondline,by=point1},
            name intersections={of=secondline and thirdline,by=point2},
            name intersections={of=thirdline and fourthline,by=point3},
            name intersections={of=fourthline and firstline,by=point4},
            ][fill={rgb, 255:red, 229; green, 229; blue, 234 }  ,fill opacity=1 ](point1)--(point2)--(point3)--(point4)--(point1);
            \node[circle,fill,inner sep=2pt] at (axis cs:2.5,0.5) {};
        \end{axis}
    \end{tikzpicture}
    \caption{Domain of convergence of the energy integral. The black dot at the center of the domain corresponds to the KZ energy spectrum.}
    \label{fig:convergenceDomain}
\end{figure}

We have seen that the most interesting exact solutions of the kinetic equations are the KZ spectra at constant energy flux. However, these solutions are only fully relevant if they satisfy the locality condition. Mathematically, this condition means that the integral must be convergent. If it is not the case, it means physically that the inertial range is not independent of the largest or smallest scales, where forcing and dissipation are expected. 
The calculation of the locality condition is highly non-trivial in this anisotropic case. It requires a careful treatment that we leave to Appendix \ref{appendix:Locality}. Note that the study of locality is still a subject of investigation \citep{Dematteis_2022}. 
In the absence of helicity, we find the following conditions 
\begin{eqnarray}
    3 < x &+& 2 y < 4, \\
    2 < x &+& y < 4.
\end{eqnarray}
We obtain a classical result for wave turbulence in the sense that the power law indices of the KZ spectra fall exactly in the middle of the convergence domain (see Figure \ref{fig:convergenceDomain}).


\section{Super-local interactions}\label{sec:SuperLocal}

In this section, we shall study the limit of local triadic interactions (in the perpendicular direction) for which the wave kinetic equations simplify significantly. From to the results found in the previous section, we know that it is mainly relevant to study the energy only. 
In the strongly anisotropic limit $k_\| \ll k_\perp$, equation (\ref{eq:EHcinetiqueLocal}) writes
\eq{}{
\pdt E_k = \sum_{s_k s_p s_q} \int T_{\bs{k p q}}^{s_k s_p s_q} \dd p_\perp \dd p_\| \dd q_\perp \dd p_\|.
}
By definition (the small parameter $\epsilon$ is absorbed in the time variable)
\eq{}{
\begin{split}
    T_{\bs{k p q}}^{s_k s_p s_q} =& \frac{ \Omega_e^2}{{2^{12}}}\frac{s_k s_p k_\| p_\|}{k_\perp^2 p_\perp^2 q_\perp^2} \left(\frac{s_q q_\perp - s_p p_\perp}{ s_k \omega_k}\right)^2 \left( s_k k_\perp + s_p p_\perp + s_q q_\perp \right)^2 \sin \alpha_q \\
    &\times E_q \left( p_\perp E_k - k_\perp E_p \right) \delta \left( \Omega_{kpq} \right) \delta_{k_\|p_\|q_\|},
\end{split}
}
is the nonlinear operator which describes the energy transfer between modes which verifies the following symmetry
\be
T_{\bs{k p q}}^{s_k s_p s_q} = - T_{\bs{p k q}}^{s_p s_k s_q}.
\ee
In the limit of super-local interactions, we can write
\eq{}{
p_\perp = k_\perp (1+\epsilon_p) \quad ; \quad q_\perp = k_\perp (1+\epsilon_q),
}
with $0 \ll \epsilon_p \ll 1$ and $0 \ll \epsilon_q \ll 1$. We can introduce an arbitrary function $f \left( k_\perp, k_\| \right)$ and integrate the kinetic equation to find
\eq{}{
\begin{split}
&\pdt  \int f \left( k_\perp, k_\| \right) E_k \dd k_\perp \dd k_\| =  \sum_{s_k s_p s_q} \int f \left( k_\perp, k_\| \right) T_{\bs{k p q}}^{s_k s_p s_q} \dd k_\perp \dd p_\perp \dd q_\perp \dd k_\| \dd p_\| \dd q_\| \\
&= \frac{1}{2} \sum_{s_k s_p s_q} \int \left[f \left( k_\perp, k_\| \right) - f \left( p_\perp, p_\| \right) \right]T_{\bs{k p q}}^{s_k s_p s_q} \dd k_\perp \dd k_\| \dd p_\perp \dd p_\| \dd q_\perp \dd p_\|. \\
\end{split}
}
Neglecting the parallel wavenumber contribution (this assumption is fully compatible with the weak cascade along the parallel direction -- see arguments based on the resonance condition), for local interactions we have
\be
f\left( p_\perp, p_\| \right) = \sum_{n=0}^{+\infty} \frac{\left(p_\perp - k_\perp \right)^n}{n!} \partial_{k_\perp}^{(n)} f \left( k_\perp, k_\| \right) = \sum_{n=0}^{+\infty} \epsilon_p^n \frac{k_\perp^n}{n!} \partial_{k_\perp}^{(n)} f \left( k_\perp, k_\| \right).
\ee
At the main order, we can write
\be
\begin{split}
    \pdt \int& f \left( k_\perp, k_\| \right) E_k \dd k_\perp \dd k_\| = \\
    &- \frac{1}{2} \partial_{k_\perp} \left( \sum_{s_k s_p s_q} \int \epsilon_p k_\perp T_{\bs{k p q}}^{s_k s_p s_q} \partial_{k_\perp} f \left( k_\perp, k_\| \right) \dd k_\perp \dd k_\| \dd p_\perp \dd p_\| \dd q_\perp \dd p_\| \right).
\end{split}
\ee
Using an integration by part, we find the relation
\be
\pdt E_k = \frac{1}{2} \partial_{k_\perp} \left( \sum_{s_k s_p s_q} \int \epsilon_p k_\perp T_{\bs{k p q}}^{s_k s_p s_q} \dd p_\perp \dd p_\| \dd q_\perp \dd p_\| \right) .
\ee
 The asymptotic form of $T_{\bs{k p q}}^{s_k s_p s_q}$ can be found by using the locality in the perpendicular direction. In particular, we find the relations
\begin{eqnarray}
k_\perp^2 p_\perp^2 q_\perp^2 &=& k_\perp^6, \\
\left(\frac{s_q q_\perp - s_p p_\perp}{ s_k \omega_k}\right)^2 &=& \frac{k_\perp^4}{\Omega_e^2 k_\|^2} \left(s_q - s_p\right), \\
\left( s_k k_\perp + s_p p_\perp + s_q q_\perp \right)^2 &=& k_\perp^2 \left( s_k  + s_p + s_q \right)^2, \\
E_q \left( p_\perp E_k - k_\perp E_p \right) &=& - \frac{\epsilon_p}{2} k_\perp^4 \partial_{k_\perp} \left( E_k/k_\perp \right)^2, \\
\sin \alpha_q &=& \sin \pi/3 = \sqrt{3}/2, \\
\delta_{\Omega_{kpq}} &=& \frac{k_\perp}{\Omega_e} \delta\left( s_k k_\| + s_p p_\| + s_q q_\| \right).
\end{eqnarray}
After simplification, we arrive at
\eq{}{
\begin{split}
    T_{\bs{k p q}}^{s_k s_p s_q} =& -\epsilon_p \frac{\sqrt{3}}{{2^{14}}} \frac{1}{\Omega_e} \frac{s_p p_\|}{s_k k_\|} k_\perp^5 \left(s_q - s_p\right)^2 \left( s_k  + s_p + s_q \right)^2 \partial_{k_\perp} \left( E_k/k_\perp \right)^2 \\  &\times \delta\left( s_k k_\| + s_p p_\| + s_q q_\| \right)
    \delta\left( k_\| + p_\| + q_\| \right).
\end{split}
}
With this form we see that the transfer will be significantly higher when $s_p =-s_q$, therefore we will only consider this type of interaction. Then, the expression of the transfer reduces to
\eq{}{
T_{\bs{k p q}}^{s_k s_p -s_p} = -\epsilon_p  \frac{\sqrt{3}}{{2^{12}}} \frac{1}{\Omega_e} \frac{s_p p_\|}{s_k k_\|} k_\perp^5 \partial_{k_\perp} \left( E_k/k_\perp \right)^2 \delta\left( s_k k_\| + s_p p_\| - s_p q_\| \right) \delta\left( k_\| + p_\| + q_\| \right).
}
The resonance condition leads to two possible combinations for the parallel wavenumbers, 
\eq{}{
\begin{split}
&k_\| + p_\| - q_\| = 0 \quad \rm{and} \quad 
k_\| + p_\| + q_\| = 0 , \\
&k_\| - p_\| + q_\| = 0 \quad \rm{and} \quad 
k_\| + p_\| + q_\| = 0  .
\end{split}
}
The solution corresponds either to $q_\| = 0$ or $p_\| = 0$, which means in particular that the strong locality assumption is not allowed for the parallel direction. The second solution cancels the transfer, therefore, we will only consider the first solution for which we have (with $p_\|=-k_\|$). We find
\eq{}{
\begin{split}
\pdt E_k &= \frac{1}{2} \partial_{k_\perp} \left( \int \epsilon_p k_\perp T_{\bs{k p q}}^{+ + -}  \dd p_\perp \dd p_\| \dd q_\perp \dd p_\| \right) \\
&= \frac{\sqrt{3}}{{2^{13}}} \frac{1}{\Omega_e} \partial_{k_\perp} \left[ k_\perp^8 \partial_{k_\perp} \left( E_k/k_\perp \right)^2 \right]  \int_{- \tilde \epsilon}^{+ \tilde \epsilon} \epsilon_p^2 \dd \epsilon_p \int_{- \tilde \epsilon}^{+ \tilde \epsilon}\dd \epsilon_q. \\
\end{split}.
}
We finally obtain the nonlinear diffusion equation 
\eq{}{
\frac{\partial E_k}{\partial t} = C \frac{\partial }{\partial k_\perp} \left[ k_\perp^8 \frac{\partial}{\partial k_\perp} \left( \frac{E_k}{k_\perp}\right)^2 \right],
}
where $C = \tilde \epsilon^4 / \left({2^{11}} \sqrt{3} \Omega_e \right)$. This equation has been derived analytically from the kinetic equations in the limit of super-local (perpendicular) interactions and when $H=0$. It gives a first interesting description of wave turbulence in IEMHD. In particular, the thermodynamic and KZ spectra are exact solutions. We can also prove that the corresponding energy flux is positive, and thus that the cascade is direct. 

It is interesting to note that a similar nonlinear diffusion equation has been obtained, in the same approximation of wave turbulence, for EMHD \citep{David_2019,Passot_2019} and rotating hydrodynamics \citep{Galtier_2020}. 
The numerical simulations of this equation reveal the existence of a $k_\perp^{-8/3}$ energy spectrum during the non-stationary phase that is steeper than the KZ spectrum. This solution has been understood as a self-similar solution of second kind (which means it cannot be predicted analytically). It is also shown that once the energy spectrum reaches the dissipative scales, a spectral bounce appears which affects the whole inertial range to finally form the expected KZ spectrum in $k_\perp^{-5/2}$.

\section{Direction of the energy cascade and Kolmogorov constant}\label{sec:Flux}

\subsection{Direct energy cascade}

\begin{figure}
    \centering

\tikzset{every picture/.style={line width=0.75pt}} 
\begin{tikzpicture}[x=0.75pt,y=0.75pt,yscale=-.67,xscale=.67]

\draw  [fill={rgb, 255:red, 155; green, 155; blue, 155 }  ,fill opacity=0.15 ] (59,75) .. controls (59,47.39) and (88.1,25) .. (124,25) .. controls (159.9,25) and (189,47.39) .. (189,75) .. controls (189,102.61) and (159.9,125) .. (124,125) .. controls (88.1,125) and (59,102.61) .. (59,75) -- cycle ;
\draw  [fill={rgb, 255:red, 155; green, 155; blue, 155 }  ,fill opacity=0.15 ] (74,75) .. controls (74,55.67) and (96.39,40) .. (124,40) .. controls (151.61,40) and (174,55.67) .. (174,75) .. controls (174,94.33) and (151.61,110) .. (124,110) .. controls (96.39,110) and (74,94.33) .. (74,75) -- cycle ;
\draw  [fill={rgb, 255:red, 155; green, 155; blue, 155 }  ,fill opacity=0.15 ] (89,75) .. controls (89,63.95) and (104.67,55) .. (124,55) .. controls (143.33,55) and (159,63.95) .. (159,75) .. controls (159,86.05) and (143.33,95) .. (124,95) .. controls (104.67,95) and (89,86.05) .. (89,75) -- cycle ;
\draw    (74,75) -- (74,106) ;
\draw    (174,75) -- (174,106) ;
\draw    (89,75) -- (89,99.83) ;
\draw    (159,75) -- (158.33,100.17) ;
\draw    (124,5) -- (124,75) ;
\draw [shift={(124,2)}, rotate = 90] [fill={rgb, 255:red, 0; green, 0; blue, 0 }  ][line width=0.08]  [draw opacity=0] (8.93,-4.29) -- (0,0) -- (8.93,4.29) -- cycle    ;
\draw    (189,160) -- (236,160) ;
\draw [shift={(239,160)}, rotate = 180] [fill={rgb, 255:red, 0; green, 0; blue, 0 }  ][line width=0.08]  [draw opacity=0] (8.93,-4.29) -- (0,0) -- (8.93,4.29) -- cycle    ;
\draw    (189,75) -- (189,222) ;
\draw    (59,75) -- (59,222) ;
\draw  [draw opacity=0] (189,222) .. controls (189,222) and (189,222) .. (189,222) .. controls (189,249.61) and (159.9,272) .. (124,272) .. controls (88.1,272) and (59,249.61) .. (59,222) -- (124,222) -- cycle ; \draw   (189,222) .. controls (189,222) and (189,222) .. (189,222) .. controls (189,249.61) and (159.9,272) .. (124,272) .. controls (88.1,272) and (59,249.61) .. (59,222) ;  
\draw    (12,160) -- (59,160) ;
\draw [shift={(9,160)}, rotate = 360] [fill={rgb, 255:red, 0; green, 0; blue, 0 }  ][line width=0.08]  [draw opacity=0] (8.93,-4.29) -- (0,0) -- (8.93,4.29) -- cycle    ;
\draw    (124,256.5) -- (124,210) ;
\draw [shift={(124,259.5)}, rotate = 270] [fill={rgb, 255:red, 0; green, 0; blue, 0 }  ][line width=0.08]  [draw opacity=0] (8.93,-4.29) -- (0,0) -- (8.93,4.29) -- cycle    ;
\draw    (201.19,229.4) -- (167.67,196.67) ;
\draw [shift={(203.33,231.5)}, rotate = 224.32] [fill={rgb, 255:red, 0; green, 0; blue, 0 }  ][line width=0.08]  [draw opacity=0] (8.93,-4.29) -- (0,0) -- (8.93,4.29) -- cycle    ;
\draw    (48.72,230.98) -- (80.33,197.33) ;
\draw [shift={(46.67,233.17)}, rotate = 313.21] [fill={rgb, 255:red, 0; green, 0; blue, 0 }  ][line width=0.08]  [draw opacity=0] (8.93,-4.29) -- (0,0) -- (8.93,4.29) -- cycle    ;

\draw (131,5.4) node [anchor=north west][inner sep=0.75pt]    {$\Pi _{\| } (\bs{k})$};
\draw (241,163.4) node [anchor=north west][inner sep=0.75pt]    {$\Pi _{\perp } (\bs{k})$};

\end{tikzpicture}
    \caption{Schematic representation of an axisymmetric flux in Fourier space. Each cylindrical shell corresponds to a specific value of $k_\perp$. In theory, they form a continuum but here their discrete nature serves as an illustration.}
    \label{fig:axisymmetricFlux}
\end{figure}
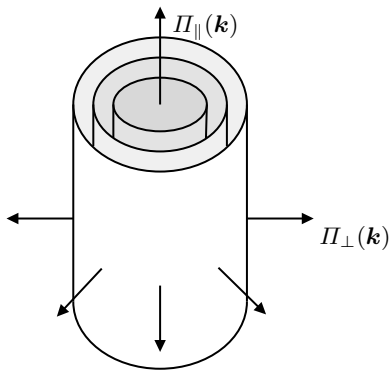

In this section, we will study the sign of the energy flux from the kinetic equations (\ref{eq:energyKZKsolution}) and prove that the cascade in the perpendicular direction is direct. In cylindrical coordinates (see figure \ref{fig:axisymmetricFlux}), we have \citep{Zakharov_1992}
\eq{}{
\frac{\partial E(\bs{k})}{\partial t}= - \bnabla \cdot \bs{\Pi} = -\frac{1}{k_\perp}\frac{\partial\left( k_\perp \Pi_\perp(\bs{k}) \right) }{\partial k_\perp} - \frac{\partial \Pi_\|(\bs{k}) }{\partial k_\|},
}
where $\bs{\Pi}$ is the energy flux vector, $\Pi_\perp$ and $\Pi_\|$ its perpendicular and parallel components (axisymmetric turbulence is assumed), respectively. Introducing the axisymmetric spectra $E_k \equiv 2 \pi k_\perp E(\bs{k})$,  $\Pi_\perp \equiv 2 \pi k_\perp \Pi_\perp(\bs{k})$ and $\Pi_\| \equiv 2 \pi k_\perp \Pi_\|(\bs{k})$, we obtain
\eq{}{
\pdt E_k = - \frac{\partial \Pi_\perp}{\partial k_\perp} - \frac{\partial \Pi_\|}{\partial k_\|} .
}

We now introduce the adimensional variables $\tilde{p}_\perp \equiv p_\perp / k_\perp$, $\tilde{q}_\perp \equiv q_\perp / k_\perp$, $\tilde{p}_\| \equiv p_\|/ k_\| $ and $\tilde{q}_\| \equiv q_\| / k_\|$. 
We seek power law solutions of the form (\ref{eq:plsol}) and then obtain
\eq{}{
\pdt E_k = \frac{\epsilon^2}{{2^{13}} \Omega_e} \left[ k_\perp^{4-2x} \left\vert k_\| \right\vert^{-2y} C_E^2 I_E(x,y) + k_\perp^{2-2\tilde{x}} \left\vert k_\| \right\vert^{-2\tilde{y}} C_H^2 I_H(\tilde{x},\tilde{y}) \right],
}
where
\eq{}{
\begin{split}
    I_E(x,y) =& \sum_{s_k s_p s_q} \int_{\Delta_\perp} s_k s_p \frac{\tilde{p}_\|}{\tilde{p}_\perp \tilde{q}_\perp^2} \left( s_q \tilde{q}_\perp - s_p \tilde{p}_\perp \right)^2 \left( s_k + s_p \tilde{p}_\perp + s_q \tilde{q}_\perp \right)^2 \sin \alpha_q \tilde{q}_\perp^{-x} \left\vert \tilde{q}_\| \right\vert^{-y} \\
    &\times \left(1-\tilde{p}_\perp^{-1-x} \left\vert \tilde{p}_\| \right\vert^{-y}\right)\left(1 - \tilde{p}_\perp^{-5+2x} \left\vert \tilde{p}_\| \right\vert^{-1+2y}\right) \\
    &\times \delta \left( s_k + s_p \frac{\tilde{p}_\|}{\tilde{p}_\perp} + s_q \frac{\tilde{q}_\|}{\tilde{q}_\perp} \right) \delta \left( 1 + \tilde{p}_\| + \tilde{q}_\| \right)  \dd \tilde{p}_\perp \dd \tilde{q}_\perp \dd \tilde{p}_\| \dd \tilde{q}_\|,
\end{split}
}
and 
\eq{}{
\begin{split}
    I_H(\tilde{x},\tilde{y}) =& \sum_{s_k s_p s_q} \int_{\Delta_\perp} s_p s_q \frac{\tilde{p}_\|}{\tilde{p}_\perp \tilde{q}_\perp^3} \left( s_q \tilde{q}_\perp - s_p \tilde{p}_\perp \right)^2 \left( s_k + s_p \tilde{p}_\perp + s_q \tilde{q}_\perp \right)^2 \sin \alpha_q \tilde{q}_\perp^{-\tilde{x}} \left\vert \tilde{q}_\| \right\vert^{-\tilde{y}} \\
    &\times \left( 1  - s_k s_p \tilde{p}_\perp^{-\tilde{x}-2} \left\vert \tilde{p}_\| \right\vert^{-\tilde{y}}\right) \left( 1 - \tilde{p}_\perp^{2\tilde{x}-3} \left\vert \tilde{p}_\| \right\vert^{2\tilde{y}-1} \right) \\
    &\times \delta \left( s_k + s_p \frac{\tilde{p}_\|}{\tilde{p}_\perp} + s_q \frac{\tilde{q}_\|}{\tilde{q}_\perp} \right)  \delta \left( 1 + \tilde{p}_\| + \tilde{q}_\| \right)  \dd \tilde{p}_\perp \dd \tilde{q}_\perp \dd \tilde{p}_\| \dd \tilde{q}_\|.
\end{split}
}
Taking the limits, corresponding to the KZ spectra, $\left( x, y, \tilde{x}, \tilde{y} \right) \rightarrow \left( 5/2, 1/2, 3/2, 1/2 \right)$, thanks to the Hospital rule, we can write
\eq{energyFluxPerpKZ}{
    \left( \Pi_\perp^\mathrm{KZ}  \atop \Pi_\|^\mathrm{KZ} \right) = \frac{\epsilon^2}{ {2^{13}} \Omega_e} \left( k_\|^{-1} \atop  k_\perp^{-1} \right) \left[ C_E^2 \left( I_\perp \atop I_\| \right)+ C_H^2 \left( J_\perp  \atop J_\| \right) \right],
}
where
\eq{Iperppara}{
\begin{split}
    \left( I_\perp \atop I_\| \right) \equiv& \sum_{s_k s_p s_q} \int_{\Delta_\perp} s_k s_p \frac{\tilde{p}_\|}{\tilde{p}_\perp \tilde{q}_\perp^{9/2} \left\vert \tilde{q}_\| \right\vert^{1/2}} \left( s_q \tilde{q}_\perp - s_p \tilde{p}_\perp \right)^2 \left( s_k + s_p \tilde{p}_\perp + s_q \tilde{q}_\perp \right)^2 \sin \alpha_q \log \left\vert \left( \tilde{p}_\perp \atop \tilde{p}_\| \right) \right\vert \\
    &\times \left(1- \tilde{p}_\perp^{-7/2} \left\vert \tilde{p}_\| \right\vert^{-1/2}\right) \delta \left( s_k + s_p \frac{\tilde{p}_\|}{\tilde{p}_\perp} + s_q \frac{\tilde{q}_\|}{\tilde{q}_\perp} \right)  \delta \left( 1 + \tilde{p}_\| + \tilde{q}_\| \right)  \dd \tilde{p}_\perp \dd \tilde{q}_\perp \dd \tilde{p}_\| \dd \tilde{q}_\|,
\end{split}
}
and
\eq{}{
\begin{split}
    \left( J_\perp \atop J_\| \right) \equiv& \sum_{s_k s_p s_q} \int_{\Delta_\perp} s_p s_q \frac{\tilde{p}_\|}{\tilde{p}_\perp \tilde{q}_\perp^{9/2} \left\vert \tilde{q}_\| \right\vert^{1/2}} \left( s_q \tilde{q}_\perp - s_p \tilde{p}_\perp \right)^2 \left( s_k + s_p \tilde{p}_\perp + s_q \tilde{q}_\perp \right)^2 \sin \alpha_q \log \left\vert \left( \tilde{p}_\perp \atop \tilde{p}_\| \right) \right\vert \\
    &\times \left(s_k - s_p \tilde{p}_\perp^{-7/2} \left\vert \tilde{p}_\| \right\vert^{-1/2}\right) \delta \left( s_k + s_p \frac{\tilde{p}_\|}{\tilde{p}_\perp} + s_q \frac{\tilde{q}_\|}{\tilde{q}_\perp} \right)  \delta \left( 1 + \tilde{p}_\| + \tilde{q}_\| \right)  \dd \tilde{p}_\perp \dd \tilde{q}_\perp \dd \tilde{p}_\| \dd \tilde{q}_\|.
\end{split}
}
Therefore, the ratio of the two fluxes is
\eq{fluxRatio}{
\frac{\Pi_\|^\mathrm{KZ}}{\Pi_\perp^\mathrm{KZ}} = \frac{k_\|}{k_\perp} \frac{C_E^2 I_\| + C_H^2 J_\|}{C_E^2 I_\perp + C_H^2 J_\perp}.
}
Since it is proportional to $k_\| / k_\perp$, we expect $\Pi_\|^\mathrm{KZ} \ll \Pi_\perp^\mathrm{KZ}$, which is in agreement with the analysis based on the resonance condition to find the direction of the cascade. 
In the absence of helicity, the ratio (\ref{eq:fluxRatio}) only depends on $k_\| / k_\perp \ll 1$ and $ I_\| / I_\perp$; we numerically find $I_\| / I_\perp \simeq 0.73$, then the previous expectation is fulfilled. 

\begin{figure}
    \centering
    \includegraphics[width=\textwidth]{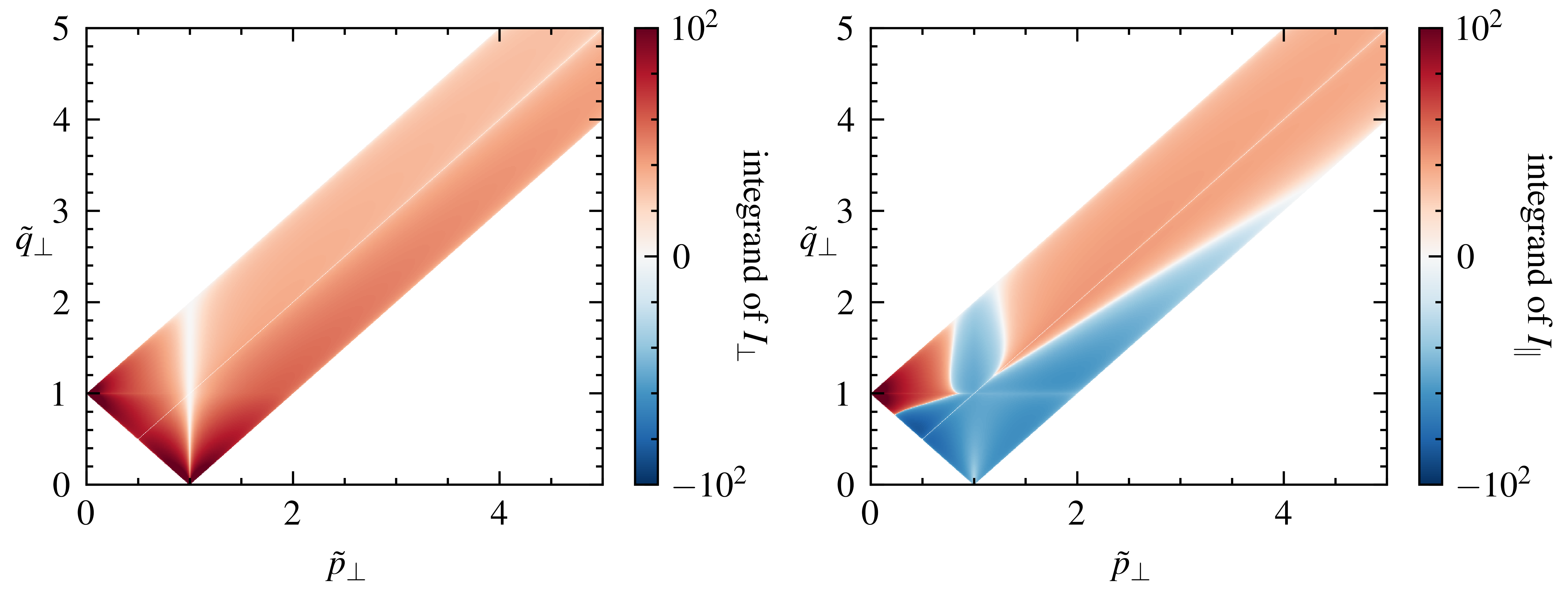}
    \caption{Left: integrand of $I_\perp$ as a function of $\tilde{p}_\perp$ and $\tilde{q}_\perp$; a positive value is always observed. Right: integrand of $I_\|$ which changes sign as a function of (small) $\tilde{p}_\perp$ and $\tilde{q}_\perp$.}
    \label{fig:fluxDifference}
\end{figure}

We can also find the sign of the energy flux and thus prove the direction of the cascade. Since the perpendicular flux is dominant, we will neglect the parallel flux and only look for the sign of $I_\perp$.
A numerical evaluation reveals a positive value, which means that $\Pi_\perp > 0$ and that the energy cascade is direct in the transverse direction. 

In figure \ref{fig:fluxDifference}, we show the sign of the integrands of $I_\perp$ and $I_\|$ obtained from a numerical evaluation of expressions (\ref{eq:Iperppara}). We see that for $I_\perp$ the integrand is always positive, while for $I_\|$ the integrand can be either positive or negative depending on the perpendicular wavenumbers (for large perpendicular wavenumbers it is always positive) but overall, after integration, the positive sign dominates in the sense that the integral $I_\| >0$. Therefore, the parallel cascade is also direct but it is composed of different contributions, with (a minority of) triadic interactions contributing to an inverse transfer.

\subsection{Kolmogorov constant}

\begin{figure}
    \centering
    \begin{tikzpicture}
\begin{axis}[small,
    legend style={at={(1,1)},anchor=north east},
    ymin=8.3,ymax=9.7,
    xmin=10,xmax=1e9,
	xmode=log, xlabel=$\xi$,
	grid=major]
\addplot[solid, line width= 1] %
	table{dataPlotCkPlasma.txt};
\addlegendentry{$C_K (\xi)$};
\addplot[dashed, line width= 0.75] %
	table{dataPlotC0Plasma.txt};
\addlegendentry{$C_K \xrightarrow[]{\xi \to \infty} 8.474$};
\end{axis}
\end{tikzpicture}
    \caption{Convergence of $C_K$ as a function of $\xi$.}
    \label{fig:ConvergenceCk}
\end{figure}
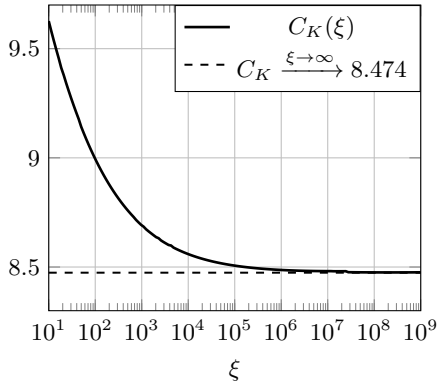

If we neglect the parallel flux and helicity, we can also obtain the expression of the Kolmogorov constant $C_K$ for which we can numerically get an estimate. To do so, we take advantage of the Dirac distributions to integrate the parallel wavenumbers. Then, since $I_\perp$ is only defined on the region $\Delta_\perp$, we introduce the change of variable $\tilde{q}_\perp \equiv \xi - \tilde{p}_\perp$ where $ \xi \in [1,+\infty[$ and $\tilde{p}_\perp \in \left[\frac{\xi-1}{2}, \frac{\xi+1}{2} \right]$ that confines the integration to this domain. One finds at a given $k_\parallel$,
\eq{}{
E_k = \sqrt{\Pi_\perp \Omega_e} C_K k_\perp^{-5/2}
\quad \rm{with} \quad
C_K = 64 \sqrt{\frac{2}{I_\perp}} \simeq 8.474 .
}
The numerical convergence of $C_K$ to this value in shown in Figure \ref{fig:ConvergenceCk}.

\section{Discussion and conclusion}\label{sec:Discussion}

In this paper, we have developed a wave turbulence theory for inertial electron MHD (i.e. for scales smaller than $d_e$) mediated by three-wave interactions between inertial whistler waves or between inertial kinetic Alfv\'en waves. 
The asymptotic wave kinetic equations are derived for the two quadratic invariants of the system, namely energy and momentum. The theory is expected to be relevant mainly for ion-electron plasmas such as the Earth's magnetosheath, the solar corona or the solar wind \citep{Milanese_2020}, but also for electron-positron plasmas \citep{Loureiro_2018}.
We show that this turbulence is mainly characterized by a direct energy cascade in the direction perpendicular to the strong applied magnetic field. The role of the second invariant, the momentum or kinetic helicity, is less important because in general there is no exact solution at constant helicity flux (except for the state of maximal helicity). By converting the exact solution (Kolmogorov-Zakharov spectrum) into unit of magnetic field, which is easier to measure in space plasma, we find a magnetic energy spectrum $E^B \sim k_\perp^{-9/2}$. It is interesting to note that this power law is steeper than that observed in the solar wind at sub-MHD scales (satisfying $k d_e < 1$) with a power law index often close to $-8/3$ \citep{Alexandrova_2012,Podesta_2013,Sahraoui2020} whereas at $kd_e$ > 1, power law indices close to $-11/3$ are observed \citep{Sahraoui_2009, Sahraoui2020} as well as $-9/2$ \citep{Sahraoui_2013} but in a narrow frequency range. The former matches the strong turbulence prediction \citep{Biskamp1999, Meyrand_2010} while the latter is in adequation with the wave turbulence one. In absence of helicity, we prove that the energy cascade is direct and numerically estimate the Kolmogorov constant using its analytical expression. We also prove that the Kolmogorov-Zakharov spectrum is in the domain of convergence, showing the relevance of the exact solution. 

An interesting point concerns the dynamics of the two-dimensional state (i.e. the slow modes for which  $k_\|=0$). We see from the kinetic equation (\ref{eq:EHcinetiqueLocal}) that the nonlinear transfer for energy and helicity decreases linearly with $k_\|$, and for the value $k_\|=0$ the transfer is exactly null. This means that the dynamics of the slow modes decouples from the three-dimensional state. 
Actually, the slow modes are not described by the wave turbulence theory which is based on the time scales separation $1/\omega \ll \tau_\mathrm{NL}$ (when $k_\| \to 0$ this inequality cannot be satisfied). 
The possibility that higher order processes, such as four-wave interactions, could lead to a coupling between two-dimensional and three-dimensional modes has been discussed in the past by \cite{Smith_1999} in the context of inertial waves in rotating hydrodynamics. Since it is a similar problem, this scenario could also be relevant here. 

In the limit of super-local (perpendicular) interactions, we derive a nonlinear diffusion equation that is similar to that found in electron MHD at scales larger than $d_e$. Interestingly, this equation is also similar to the case of inertial wave turbulence (fast rotating hydrodynamic turbulence). In fact, the link is deeper than that since the two problems share the same kinetic equations (within a factor) with the same dispersion relation (within a factor). 
This connection is due to a strong asymmetry imposed by a mean magnetic field on the one hand, and by the axis rotation on the other hand. It is also due to the helical nature of the waves.
This reinforces the bridge between plasma physics and fluid mechanics (see also \cite{Galtier_2020}) and suggests that laboratory experiments \citep{Yarom_2014, Monsalve_2020} can help to better understand space plasma physics at a scale still difficult to detect by current spacecraft.

%

\appendix
\section{Detailed conservation of  energy}\label{appendix:Conservation}

We recall the relations $u_\|= d_e \nabla^2_\perp \psi$ and ${\bf u_\perp}= d_e (- \partial_y b_\| \hat{\bs{e}}_x + \partial_x b_\| \hat{\bs{e}}_y)$ which allow us, in Fourier space, to obtain the expressions of the energy density respectively in the directions parallel and perpendicular to the mean magnetic field: $\vert u_{\|,k}\vert^2= d_e^2 k_\perp^4 \vert \psi_k \vert^2$ and $\vert u_{\perp,k}\vert^2= d_e^2 k_\perp^2 \vert b_k\vert^2$ (with $b_k \equiv {b_\|}_k$). 
From the equations describing the temporal evolution of $\psi$ and $b_\|$ in Fourier space, we obtain the evolution of the energy density 
\begin{eqnarray}
\pdt \vert u_{\|,k}\vert^2 - i d_e^2 \Omega_e k_\| k_\perp^2 b_k \psi_k + c.c. &=& d_e^3 \int_{\mathbb{R}^6}
\begin{aligned}[t]
            & \sin \alpha_k k_\perp^2 p_\perp q_\perp \left(q_\perp^2 \psi_q b_p  - p_\perp^2 \psi_p b_q \right) \\
            &\times \psi_k {\delta_{kpq}} \dd\bs{p} \dd\bs{q} + c.c. ,
        \end{aligned}\\
\pdt \vert u_{\perp,k}\vert^2 + i d_e^2 \Omega_e k_\| k_\perp^2 b_k \psi_k + c.c. &=& d_e^3 \int_{\mathbb{R}^6} 
\begin{aligned}[t]
            &\sin \alpha_k p_\perp q_\perp \left(q_\perp^2 - p_\perp^2 \right) b_k b_p b_q \\
            &\times {\delta_{kpq}} \dd\bs{p} \dd\bs{q} + c.c. ,
        \end{aligned}
\end{eqnarray}
where we have used the relation $\hat{\bs{e}}_\| \cdot \left(\hat{\bs{e}}_{p_\perp} \times \hat{\bs{e}}_{q_\perp} \right) = \sin \alpha_k$ and $c.c.$ denotes the complex conjugate. Parallel $E_\|^u$ and perpendicular $E_\perp^u$ energies being the sum of these quantities over the all wavenumbers, we find 
\begin{eqnarray}
\pdt E_\|^u - i d_e^2 \Omega_e \int_{\mathbb{R}^3} k_\| k_\perp^2  b_k \psi_k \dd\bs{k} + {c.c.} &=& d_e^3 \int_{\mathbb{R}^9} S_\|^u \left( k_\perp, p_\perp, q_\perp \right) {\delta_{kpq}} \dd\bs{k} \dd\bs{p} \dd\bs{q} + c.c., \\
\pdt E_\perp^u + i d_e^2 \Omega_e \int_{\mathbb{R}^3} k_\| k_\perp^2  b_k \psi_k \dd\bs{k} + {c.c.} &=& d_e^3 \int_{\mathbb{R}^9} S_\perp^u \left( k_\perp, p_\perp, q_\perp \right) {\delta_{kpq}} \dd\bs{k} \dd\bs{p} \dd\bs{q} + {c.c.},
\end{eqnarray}
with $S_\|^u \left( k_\perp, p_\perp, q_\perp \right)$ and $S_\perp^u \left( k_\perp, p_\perp, q_\perp \right)$ the nonlinear interaction coefficient defined as
\begin{eqnarray}
S_\|^u \left( k_\perp, p_\perp, q_\perp \right) &\equiv& \sin \alpha_k k_\perp^2 p_\perp q_\perp \psi_k \left(q_\perp^2 \psi_q b_p  - p_\perp^2 \psi_p b_q \right), \\
S_\perp^u \left( k_\perp, p_\perp, q_\perp \right) &\equiv& \sin \alpha_k p_\perp q_\perp \left(q_\perp^2 - p_\perp^2 \right) b_k b_p b_q.
\end{eqnarray}
The remarkable property is that the nonlinear contributions are both conserved over time since $S_\|^u \left( k_\perp, p_\perp, q_\perp \right)$ and $S_\perp^u \left( k_\perp, p_\perp, q_\perp \right)$ verify the following relations
\begin{eqnarray}
S_\|^u \left( k_\perp, p_\perp, q_\perp \right) + S_\|^u \left( q_\perp, k_\perp, q_\perp \right)+ S_\|^u \left( p_\perp, q_\perp, k_\perp \right) = 0, \\
S_\perp^u \left( k_\perp, p_\perp, q_\perp \right) + S_\perp^u \left( q_\perp, k_\perp, q_\perp \right)+ S_\perp^u \left( p_\perp, q_\perp, k_\perp \right) = 0.
\end{eqnarray}
Then, the parallel and perpendicular components of the energy are conserved individually at the nonlinear level. The exchanges between the two are only done at the linear level.

\section{Derivation of the wave kinetic equations}\label{appendix:Moments}

We start from (\ref{eq:kineticFundamental}) and write successively equations for the second- and third-order moments,
\eq{B1eq}{
\begin{split}
\pdt \left\langle a_{\bs{k}}^{s_k} a_{\bs{k'}}^{s_k'} \right\rangle =& 
\epsilon \sum_{s_p s_q} \int_{\mathbb{R}^6} L_{\bs{kpq}}^{s_k s_p s_q} \left\langle  a_{\bs{k'}}^{s_k'} a_{\bs{p}}^{s_p} a_{\bs{q}}^{s_q} \right\rangle \mathrm{e}^{i \Omega_{pq}^k t} \delta_{pq}^k \dd\bs{p} \dd\bs{q}  \\
&+  \epsilon \sum_{s_p s_q} \int_{\mathbb{R}^6} L_{\bs{k'pq}}^{s_k' s_p s_q} \left\langle a_{\bs{k}}^{s_k} a_{\bs{p}}^{s_p} a_{\bs{q}}^{s_q} \right\rangle \mathrm{e}^{i \Omega_{pq}^{k'}t} \delta_{pq}^{k'} \dd\bs{p} \dd\bs{q},
\end{split}
}
and
\eq{}{
\begin{split}
\pdt \left\langle a_{\bs{k}}^{s_k} a_{\bs{k'}}^{s_k'} a_{\bs{k''}}^{s_k''} \right\rangle =&
\epsilon \sum_{s_p s_q} \int_{\mathbb{R}^6} L_{\bs{kpq}}^{s_k s_p s_q} \left\langle  a_{\bs{k}'}^{s_k'} a_{\bs{k}''}^{s_k''} a_{\bs{p}}^{s_p} a_{\bs{q}}^{s_q} \right\rangle \mathrm{e}^{i \Omega_{pq}^k t} \delta_{pq}^{k} \dd\bs{p} \dd\bs{q} \\ 
&+ 
\epsilon \sum_{s_p s_q} \int_{\mathbb{R}^6} L_{\bs{k'pq}}^{s_k' s_p s_q} \left\langle  a_{\bs{k}}^{s_k} a_{\bs{k}''}^{s_k''} a_{\bs{p}}^{s_p} a_{\bs{q}}^{s_q} \right\rangle \mathrm{e}^{i \Omega_{pq}^{k'}t} \delta_{pq}^{k'} \dd\bs{p} \dd\bs{q} \\ 
&+
\epsilon \sum_{s_p s_q} \int_{\mathbb{R}^6} L_{\bs{k''pq}}^{s_k'' s_p s_q} \left\langle  a_{\bs{k}}^{s_k} a_{\bs{k}'}^{s_k'} a_{\bs{p}}^{s_p} a_{\bs{q}}^{s_q} \right\rangle \mathrm{e}^{i \Omega_{pq}^{k''}t} \delta_{pq}^{k''}  \dd\bs{p} \dd\bs{q}.
\end{split}
}
A natural closure arises for times asymptotically large compare to the linear wave time scale (see e.g. \cite{Newell_2001, Nazarenko_2011, Newell_2011}). 
An important aspect is the uniformity of the development which was discussed first by \cite{Benney1966}. 
In this case, the fourth-order moment does not contribute at large time and, therefore, the nonlinear regeneration of third-order moments depends essentially on products of second-order moments
\eq{}{
\left\langle  a_{\bs{k}'}^{s_k'} a_{\bs{k}''}^{s_k''} a_{\bs{p}}^{s_p} a_{\bs{q}}^{s_q} \right\rangle =
\left\langle a_{\bs{p}}^{s_p} a_{\bs{q}}^{s_q} \right\rangle \left\langle a_{\bs{k}'}^{s_k'} a_{\bs{k}''}^{s_k''} \right\rangle
+
\left\langle a_{\bs{p}}^{s_p} a_{\bs{k}'}^{s_k'} \right\rangle \left\langle a_{\bs{q}}^{s_q} a_{\bs{k}''}^{s_k''} \right\rangle
+
\left\langle a_{\bs{p}}^{s_p} a_{\bs{k}''}^{s_k''} \right\rangle \left\langle a_{\bs{q}}^{s_q} a_{\bs{k}'}^{s_k'} \right\rangle.
}
Thanks to the integration on the dummy variables $\bs{p}$ and $\bs{q}$, to their symmetry and the symmetry between the polarizations $s_p$ and $s_q$, we make the following simplification in advance
\eq{}{
\left\langle  a_{\bs{k}'}^{s_k'} a_{\bs{k}''}^{s_k''} a_{\bs{p}}^{s_p} a_{\bs{q}}^{s_q} \right\rangle =
\left\langle a_{\bs{p}}^{s_p} a_{\bs{q}}^{s_q} \right\rangle \left\langle a_{\bs{k}'}^{s_k'} a_{\bs{k}''}^{s_k''} \right\rangle
+
2 \left\langle a_{\bs{p}}^{s_p} a_{\bs{k}'}^{s_k'} \right\rangle \left\langle a_{\bs{q}}^{s_q} a_{\bs{k}''}^{s_k''} \right\rangle,
}
and also introduce the spectral energy density $e^{s_k'}\left( \bs{k}'\right) = e_{\bs{k}'}^{s_k'}$ such as
\eq{}{
\left\langle a_{\bs{k}}^{s_k} a_{\bs{k'}}^{s_k'} \right\rangle = e_{\bs{k}'}^{s_k'} \delta_{kk'}\delta^{s_k}_{s_k'},
}
where $\delta_{kk'} = \delta\left(\bs{k} + \bs{k}'\right)$ and $\delta^{s_k}_{s_k'} = \delta\left(s_k - s_k'\right)$. The last delta condition ensures that the contribution is non-negligible 
over long times. We then write 
\begin{eqnarray}
\left\langle a_{\bs{k}'}^{s_k'} a_{\bs{k}''}^{s_k''} a_{\bs{p}}^{s_p} a_{\bs{q}}^{s_q} \right\rangle
&=&
e_{\bs{p}}^{s_p} \delta_{pq}\delta^{s_p}_{s_q} e_{\bs{k'}}^{s_k'} \delta_{k'k''}\delta^{s_k'}_{s_k''}
+
2 e_{\bs{p}}^{s_p} \delta_{pk'}\delta^{s_p}_{s_k'} e_{\bs{q}}^{s_q} \delta_{qk''}\delta^{s_q}_{s_k''}, \\
\left\langle  a_{\bs{k}}^{s_k} a_{\bs{k}''}^{s_k''} a_{\bs{p}}^{s_p} a_{\bs{q}}^{s_q} \right\rangle
&=&
e_{\bs{p}}^{s_p} \delta_{pq}\delta_{s_q}^{s_p} e_{\bs{k}}^{s_k} \delta_{kk''}\delta^{s_k}_{s_k''}
+
2 e_{\bs{p}}^{s_p} \delta_{pk}\delta^{s_p}_{s_k} e_{\bs{q}}^{s_q} \delta_{qk''}\delta^{s_q}_{s_k''}, \\
\left\langle  a_{\bs{k}}^{s_k} a_{\bs{k}'}^{s_k'} a_{\bs{p}}^{s_p} a_{\bs{q}}^{s_q} \right\rangle
&=&
e_{\bs{p}}^{s_p} \delta_{pq}\delta^{s_p}_{s_q} e_{\bs{k}}^{s_k} \delta_{kk'}\delta^{s_k}_{s_k'}
+
2 e_{\bs{p}}^{s_p} \delta_{pk}\delta^{s_p}_{s_k} e_{\bs{q}}^{s_q} \delta_{qk'}\delta^{s_q}_{s_k'}.
\end{eqnarray}
We note that, on the one hand, the $\delta_{pq}$ imposes $\bs{p}=-\bs{q}$ and, on the other hand, the $\delta_{pq}^{k}$ imposes $\bs{k}=\bs{p}+\bs{q}$. Thus, these two conditions lead to $\bs{k}=\bs{0}$. Since $L_{\bs{0pq}}^{s_k s_p s_q}= 0$, the first term on the right side hand side is zero and we get
\eq{}{
\begin{split}
\pdt \left\langle a_{\bs{k}}^{s_k} a_{\bs{k'}}^{s_k'} a_{\bs{k''}}^{s_k''} \right\rangle &=
2 \epsilon \sum_{s_p s_q} \int_{\mathbb{R}^6} L_{\bs{-kpq}}^{s_k s_p s_q} e_{\bs{p}}^{s_p} e_{\bs{q}}^{s_q} \mathrm{e}^{i \Omega_{pq}^k t} \delta^{s_p}_{s_k'} \delta^{s_q}_{s_k''} \delta_{pk'} \delta_{qk''} \delta_{pq}^k \dd\bs{p} \dd\bs{q} \\ 
&\quad + 
2 \epsilon \sum_{s_p s_q} \int_{\mathbb{R}^6} L_{\bs{-k'pq}}^{s_k' s_p s_q} e_{\bs{p}}^{s_p} e_{\bs{q}}^{s_q} \mathrm{e}^{i \Omega_{pq}^{k'}t} \delta^{s_p}_{s_k} \delta^{s_q}_{s_k''} \delta_{pk} \delta_{qk''} \delta_{pq}^{k'} \dd\bs{p} \dd\bs{q} \\ 
&\quad +
2 \epsilon \sum_{s_p s_q} \int_{\mathbb{R}^6} L_{\bs{-k''pq}}^{s_k'' s_p s_q} e_{\bs{p}}^{s_p} e_{\bs{q}}^{s_q} \mathrm{e}^{i \Omega_{pq}^{k''}t} \delta^{s_p}_{s_k} \delta^{s_q}_{s_k'} \delta_{pk} \delta_{qk'} \delta_{pq}^{k''}  \dd\bs{p} \dd\bs{q}.
\end{split}
}
After integration and summation over the polarizations, we obtain
\eq{}{
\begin{split}
& \pdt \left\langle a_{\bs{k}}^{s_k} a_{\bs{k'}}^{s_k'} a_{\bs{k''}}^{s_k''} \right\rangle =
2 \epsilon \mathrm{e}^{i \Omega_{kk'k''} t}  \delta_{kk'k''} \\
&\times \left( L_{\bs{-k-k'-k''}}^{s_k s_k' s_k''} e_{\bs{-k'}}^{s_k'} e_{\bs{-k''}}^{s_k''} 
+
L_{\bs{-k'-k-k''}}^{s_k' s_k s_k''} e_{\bs{-k}}^{s_k} e_{\bs{-k''}}^{s_k''} +
L_{\bs{-k''-k-k'}}^{s_k'' s_k s_k'} e_{\bs{-k}}^{s_k} e_{\bs{-k'}}^{s_k'} \right),
\end{split}
}
where $\Omega_{kk'k''} = s_k \omega_k + s_k' \omega_k' + s_k'' \omega_k'' $. 
Further simplifications can be made. Firstly, the energy density tensor describes an homogeneous turbulence then $e_{\bs{-k}}^{s_k}=e_{\bs{k}}^{s_k}$. Secondly, the interaction coefficient has the following symmetry $L_{\bs{-k-k'-k''}}^{s_k s_k' s_k''}=L_{\bs{kk'k''}}^{s_k s_k' s_k''}$.
Thirdly, we introduce $L_{\bs{kpq}}^{s_k s_p s_q} \equiv \left(s_q q_\perp - s_p p_\perp \right) M_{\bs{kpq}}^{s_k s_p s_q}$ which is convenient for the calculations. We obtain
\eq{}{
\begin{split}
\pdt \left\langle a_{\bs{k}}^{s_k} a_{\bs{k'}}^{s_k'} a_{\bs{k''}}^{s_k''} \right\rangle
= 2 \epsilon \mathrm{e}^{i \Omega_{k k'k''} t}  \delta_{kk'k''} &\left[ \left( s_k''k_\perp'' - s_k' k_\perp'\right) M_{\bs{kk'k''}}^{s_k s_k' s_k''} e_{\bs{k'}}^{s_k'} e_{\bs{k''}}^{s_k''} \right. \\
&\left. + \left( s_k'' k_\perp'' - s_k k_\perp \right) M_{\bs{k'kk''}}^{s_k' s_k s_k''} e_{\bs{k}}^{s_k} e_{\bs{k''}}^{s_k''} \right.\\
&\left. + \left(s_k' k_\perp' - s_k k_\perp \right) M_{\bs{k''kk'}}^{s_k'' s_k s_k'} e_{\bs{k}}^{s_k} e_{\bs{k'}}^{s_k'}\right].
\end{split}
}
We observe that $M_{\bs{kk'k''}}^{s_k s_k' s_k''}=M_{\bs{k''kk'}}^{s_k'' s_k s_k'}=-M_{\bs{k'kk''}}^{s_k' s_k s_k''}$ thus the previous expression can be simplified
\eq{}{
\begin{split}
\pdt \left\langle a_{\bs{k}}^{s_k} a_{\bs{k'}}^{s_k'} a_{\bs{k''}}^{s_k''} \right\rangle
=
2 \epsilon \mathrm{e}^{i \Omega_{k k'k''} t}  \delta_{kk'k''}  M_{\bs{k k' k''}}^{s_k s_k' s_k''} &\left[ \left(s_k''k_\perp'' - s_k' k_\perp'\right) e_{\bs{k'}}^{s_k'} e_{\bs{k''}}^{s_k''} \right. \\
&\left. - \left(s_k''k_\perp'' - s_k k_\perp \right) e_{\bs{k}}^{s_k} e_{\bs{k''}}^{s_k''} \right. \\
&\left. + \left(s_k'k_\perp' - s_k k_\perp \right) e_{\bs{k}}^{s_k} e_{\bs{k'}}^{s_k'} \right].
\end{split}
}
We note that $s_k'' k'' - s_k' k' = s_k'' k'' - s_k k + s_k k - s_k' k'$, and thus
\eq{}{
\begin{split}
\pdt \left\langle a_{\bs{k}}^{s_k} a_{\bs{k'}}^{s_k'} a_{\bs{k''}}^{s_k''} \right\rangle
=
2 \epsilon \mathrm{e}^{i \Omega_{k k'k''} t}  \delta_{kk'k''}  M_{\bs{kk'k''}}^{s_k s_k' s_k''} &\left[ \left(s_k''k_\perp'' - s_k k_\perp\right) \left( e_{\bs{k'}}^{s_k'} e_{\bs{k''}}^{s_k''} - e_{\bs{k}}^{s_k} e_{\bs{k''}}^{s_k''} \right) \right. \\
&\left. + \left(s_k'k_\perp' - s_k k_\perp\right) \left( e_{\bs{k}}^{s_k} e_{\bs{k'}}^{s_k'} - e_{\bs{k'}}^{s_k'} e_{\bs{k''}}^{s_k''} \right) \right].
\end{split}
}
After integration over time, one has 
\eq{B14eq}{
\begin{split}
\left\langle a_{\bs{k}}^{s_k} a_{\bs{k'}}^{s_k'} a_{\bs{k''}}^{s_k''} \right\rangle
=
2 \epsilon \Delta \left(\Omega_{k k'k''}\right) \delta_{kk'k''}  M_{\bs{kk'k''}}^{s_k s_k' s_k''} &\left[
\left(s_k k_\perp - s_k''k_\perp''\right) e_{\bs{k''}}^{s_k''} \left( e_{\bs{k}}^{s_k} - e_{\bs{k'}}^{s_k'}  \right) \right. \\
&\left. +
\left(s_k' k_\perp' - s_k k_\perp\right) e_{\bs{k'}}^{s_k'} \left( e_{\bs{k}}^{s_k} - e_{\bs{k''}}^{s_k''}\right) \right],
\end{split}
}
with
\eq{}{
\Delta \left(x\right) = \int_{0}^{t\gg 1/\omega} \mathrm{e}^{i x \tau} \dd \tau = \frac{\mathrm{e}^{i x t}-1}{i x}.
}
Now, we can introduce expression (\ref{eq:B14eq}) for the third-order moment into equation (\ref{eq:B1eq})
\eq{}{
\begin{split}
\pdt \left\langle a_{\bs{k}}^{s_k} a_{\bs{k'}}^{s_k'} \right\rangle &= \pdt e_{\bs{k}}^{s_k} \delta_{kk'}\delta^{s_k}_{s_k'} \\
&= \epsilon \sum_{s_p s_q} \int_{\mathbb{R}^6} \left( L_{\bs{-kpq}}^{s_k s_p s_q} \left\langle a_{\bs{k'}}^{s_k'} a_{\bs{p}}^{s_p} a_{\bs{q}}^{s_q} \right\rangle \mathrm{e}^{i \Omega_{pq}^k t} \delta_{pq}^k \right. \\
&\left. \hspace{60pt} + L_{\bs{-k'pq}}^{s_k' s_p s_q} \left\langle a_{\bs{k}}^{s_k} a_{\bs{p}}^{s_p} a_{\bs{q}}^{s_q} \right\rangle \mathrm{e}^{i \Omega_{pq}^{k'}t} \delta_{pq}^{k'} \right) \delta^{s_k}_{s_k'} \delta_{kk'} \dd\bs{p} \dd\bs{q} \\
&= I_1 + I_2,
\end{split}
}
where $I_1$ and $I_2$ are the two integrals involving the interaction coefficients $L_{\bs{-kpq}}^{s_k s_p s_q}$ and $L_{\bs{-k'pq}}^{s_k' s_p s_q}$, respectively. Expressing $L_{\bs{kpq}}^{s_k s_p s_q}$ as a function of $M_{\bs{kpq}}^{s_k s_p s_q}$, the first integral becomes
\eq{}{
\begin{split}
I_1 =
2 \epsilon^2 \sum_{s_p s_q} \int_{\mathbb{R}^6} & \left( s_q q_\perp - s_p p_\perp \right) \left\vert M_{\bs{-kpq}}^{s_k s_p s_q} \right\vert^2  \Delta \left(\Omega_{-kpq}\right) \mathrm{e}^{i \Omega^k_{pq} t} \delta_{pq}^k \\
& \times \left[ \left(s_k k_\perp - s_q q_\perp \right) e_{\bs{q}}^{s_q} \left( e_{\bs{k}}^{s_k} - e_{\bs{p}}^{s_p}  \right) + 
\left(s_pp_\perp - s_k k_\perp\right) e_{\bs{p}}^{s_p} \left( e_{\bs{k}}^{s_k} - e_{\bs{q}}^{s_q}\right) \right] \dd\bs{p} \dd\bs{q}.
\end{split}
}
We note that $\Delta \left(\Omega_{-kpq}\right) \mathrm{e}^{i \Omega^k_{pq} t} = \Delta \left(\Omega_{pq}^k\right)$. The long time behavior is given by the Riemann-Lebesgue lemma
\eq{}{
\Delta(x) \xrightarrow{t \to \infty}  \pi \delta(x) + i \mathcal{P}\left( \frac{1}{x} \right).
}
After a last change of variable, we find $\left(\bs{p},\bs{q}\right)\to\left(-\bs{p},-\bs{q}\right)$
\eq{}{
\begin{split}
I_1 = &
2 \epsilon^2 \sum_{s_p s_q} \int_{\mathbb{R}^6} \left[
\left(s_k k_\perp - s_q q_\perp \right) e_{\bs{q}}^{s_q} \left( e_{\bs{k}}^{s_k} - e_{\bs{p}}^{s_p}  \right) + \left(s_pp_\perp - s_k k_\perp\right) e_{\bs{p}}^{s_p} \left( e_{\bs{k}}^{s_k} - e_{\bs{q}}^{s_q}\right) \right] \\
&\times \left( s_q q_\perp - s_p p_\perp \right) \left\vert M_{\bs{kpq}}^{s_k s_p s_q} \right\vert^2 \left[ \pi \delta\left(\Omega_{kpq} \right) + i \mathcal{P}\left( \frac{1}{\Omega_{kpq}} \right) \right] \delta_{kpq} \dd\bs{p} \dd\bs{q}.
\end{split}
}
The same manipulation with $I_2$ without performing the change of variable leads to
\eq{}{
\begin{split}
I_2 = &
2 \epsilon^2 \sum_{s_p s_q} \int_{\mathbb{R}^6} \left[
\left(s_k k_\perp - s_q q_\perp \right) e_{\bs{q}}^{s_q} \left( e_{\bs{k}}^{s_k} - e_{\bs{p}}^{s_p}  \right) + \left(s_pp_\perp - s_k k_\perp\right) e_{\bs{p}}^{s_p} \left( e_{\bs{k}}^{s_k} - e_{\bs{q}}^{s_q}\right) \right] \\
&\times \left( s_q q_\perp - s_p p_\perp \right) \left\vert M_{\bs{kpq}}^{s_k s_p s_q} \right\vert^2 \left[ \pi \delta\left(\Omega_{kpq} \right) - i \mathcal{P}\left( \frac{1}{\Omega_{kpq}} \right) \right] \delta_{kpq} \dd\bs{p} \dd\bs{q},
\end{split}
}
and the sum of these two integrals gives
\eq{}{
\begin{split}
\pdt e_{\bs{k}}^{s_k} = &
4 \pi \epsilon^2  \sum_{s_p s_q} \int_{\mathbb{R}^6} \left[
\left(s_k k_\perp - s_q q_\perp \right) e_{\bs{q}}^{s_q} \left( e_{\bs{k}}^{s_k} - e_{\bs{p}}^{s_p}  \right) + \left(s_p p_\perp + s_k k_\perp\right) e_{\bs{p}}^{s_p} \left( e_{\bs{k}}^{s_k} - e_{\bs{q}}^{s_q}\right) \right] \\
&\times \left( s_q q_\perp - s_p p_\perp \right) \left\vert M_{\bs{kpq}}^{s_k s_p s_q} \right\vert^2  \delta\left(\Omega_{kpq} \right) \delta_{kpq} \dd\bs{p} \dd\bs{q}.
\end{split}
}
Using the symmetries of the resonant conditions, we have
\eq{}{
\begin{split}
    \pdt e_{\bs{k}}^{s_k} =
& \frac{\pi \epsilon^2 }{16} \sum_{s_p s_q} \int_{\mathbb{R}^6} \frac{\left( s_q q_\perp - s_p p_\perp \right)^2}{s_k \omega_k} \frac{\sin^2 \alpha_k}{k_\perp^2} \left( s_k k_\perp + s_p p_\perp + s_q q_\perp \right)^2 \\
& \times \left[
s_p \omega_p e_{\bs{q}}^{s_q} \left( e_{\bs{k}}^{s_k} - e_{\bs{p}}^{s_p}  \right)
+
s_q \omega_q e_{\bs{p}}^{s_p} \left( e_{\bs{k}}^{s_k} - e_{\bs{q}}^{s_q}\right) \right] \delta\left(\Omega_{kpq} \right) \delta_{kpq} \dd\bs{p} \dd\bs{q}.
\end{split}
}
The $\delta\left(\Omega_{kpq} \right)$ allows us to finally rewrite the term in the second line as follow
\eq{}{
\begin{split}
\pdt e_{\bs{k}}^{s_k} =
\frac{\pi \epsilon^2}{16} \sum_{s_p s_q} \int_{\mathbb{R}^6}& 
\frac{1}{s_k \omega_k}
\left\vert L_{\bs{kpq}}^{s_k s_p s_q} \right\vert^2 \left( s_k \omega_k e_{\bs{p}}^{s_p} e_{\bs{q}}^{s_q} + s_p \omega_p e_{\bs{k}}^{s_k} e_{\bs{q}}^{s_q} + s_q \omega_q e_{\bs{k}}^{s_k} e_{\bs{p}}^{s_p}
\right) \\
&\times \delta\left(\Omega_{kpq} \right) \delta_{kpq} \dd\bs{p} \dd\bs{q}.
\end{split}
}
These are the kinetic equations for IEMHD wave turbulence.

\section{Locality criteria}\label{appendix:Locality}

\begin{figure}
    \centering

\tikzset{every picture/.style={line width=0.75pt}} 

\begin{tikzpicture}[x=0.75pt,y=0.75pt,yscale=-1,xscale=1]

\draw  [color={rgb, 255:red, 0; green, 0; blue, 0 }  ,draw opacity=0 ][fill={rgb, 255:red, 229; green, 229; blue, 234 }  ,fill opacity=1 ] (50.43,257.14) -- (140.04,167.54) -- (182.46,209.96) -- (92.86,299.57) -- cycle ;
\draw [color={rgb, 255:red, 199; green, 199; blue, 204 }  ,draw opacity=1 ]   (50.43,257.14) -- (92.86,299.57) ;
\draw [color={rgb, 255:red, 199; green, 199; blue, 204 }  ,draw opacity=1 ]   (92.86,299.57) -- (160.32,232.11) ;
\draw [color={rgb, 255:red, 199; green, 199; blue, 204 }  ,draw opacity=1 ]   (50.43,257.14) -- (117.89,189.68) ;
\draw [color={rgb, 255:red, 199; green, 199; blue, 204 }  ,draw opacity=1 ] [dash pattern={on 4.5pt off 4.5pt}]  (117.89,189.68) -- (140.04,167.54) ;
\draw [color={rgb, 255:red, 199; green, 199; blue, 204 }  ,draw opacity=1 ] [dash pattern={on 4.5pt off 4.5pt}]  (160.32,232.11) -- (182.46,209.96) ;
\draw    (50,300) -- (50.98,153) ;
\draw [shift={(51,150)}, rotate = 90.38] [fill={rgb, 255:red, 0; green, 0; blue, 0 }  ][line width=0.08]  [draw opacity=0] (8.93,-4.29) -- (0,0) -- (8.93,4.29) -- cycle    ;
\draw    (50,300) -- (197,300) ;
\draw [shift={(200,300)}, rotate = 180] [fill={rgb, 255:red, 0; green, 0; blue, 0 }  ][line width=0.08]  [draw opacity=0] (8.93,-4.29) -- (0,0) -- (8.93,4.29) -- cycle    ;

\draw (178,303.4) node [anchor=north west][inner sep=0.75pt]    {$\tilde{p}_{\perp }$};
\draw (30,153.4) node [anchor=north west][inner sep=0.75pt]    {$\tilde{q}_{\perp }$};
\draw (86.86,302.97) node [anchor=north west][inner sep=0.75pt]    {$1$};
\draw (37.43,249.54) node [anchor=north west][inner sep=0.75pt]    {$1$};
\draw (36,293.4) node [anchor=north west][inner sep=0.75pt]    {$0$};
\draw (86,275.4) node [anchor=north west][inner sep=0.75pt]    {A};
\draw (59,249.4) node [anchor=north west][inner sep=0.75pt]    {B};
\draw (147,187.4) node [anchor=north west][inner sep=0.75pt]    {C};

\end{tikzpicture}
    \caption{The kinetic equations are integrated on a domain verifying $\bs{k} + \bs{p} + \bs{q} = \bs{0}$. The grey strip corresponds to this domain for the adimensional perpendicular wavevectors. A, B and C (at infinity) are the non-local regions where the convergence of the integrals must be checked.}
    \label{fig:convergenceCriteria}
\end{figure}
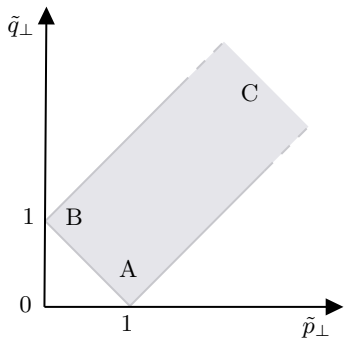

The objective of this section is to find the locality domain of the power law solutions at constant energy flux and (for simplicity) in absence of helicity. In other words, we want to check if the contribution of non-local interactions are not dominant. There are three areas (regions A, B and C in figure \ref{fig:convergenceCriteria}) for which the interactions are non-local. 
To do this, it is convenient to introduce the adimensional wavenumbers $\tilde{p}_\perp \equiv p_\perp/k_\perp$, $\tilde{p}_\| \equiv p_\|/k_\|$, $\tilde{q}_\perp \equiv q_\perp/k_\perp$ and $\tilde{q}_\| \equiv q_\|/k_\|$.
We obtain ($H=0$):
\eq{}{
\begin{split}
    \pdt E_k =& \frac{\epsilon^2 C_E^2}{{2^{12}} \Omega_e} k_\perp^{4-2x} k_\|^{-2y} \sum_{s_k s_p s_q} \int_{\Delta_\perp}  s_k s_p \frac{\tilde{p}_\|}{\tilde{p}_\perp \tilde{q}_\perp^2} \left(s_q\tilde{q}_\perp - s_p \tilde{p}_\perp  \right)^2  \left(s_k + s_p \tilde{p}_\perp + s_q \tilde{q}_\perp  \right)^2 \sin \alpha_q \\
    &\times \tilde{q}_\perp^{-x} \tilde{q}_\|^{-y} \left(1 - \tilde{p}_\perp^{-1-x} \tilde{p}_\|^{-y} \right) \delta \left(s_k + s_p \frac{\tilde{p}_\|}{\tilde{p}_\perp} + s_q \frac{\tilde{q}_\|}{\tilde{q}_\perp} \right) \delta \left(1 + \tilde{p}_\| + \tilde{q}_\| \right) \dd \tilde{p}_\perp \dd \tilde{q}_\perp \dd \tilde{p}_\| \dd \tilde{q}_\|.
\end{split}
}
This expression can be integrated in the parallel directions. We recall the following property
\eq{}{
\int_\mathbb{R} f(x) \delta \left( g(x) \right) \dd x = \sum_i \frac{f \left( x_i \right) }{\left| g' \left( x_i \right) \right|},\, \text{such as}\, g \left( x_i \right)=0.
}
Then, we have
\begin{eqnarray}
\delta \left(1 + \tilde{p}_\| + \tilde{q}_\| \right) &\longrightarrow& \tilde{q}_\| = -1 - \tilde{p}_\|, \\
\delta \left(s_k + s_p \frac{\tilde{p}_\|}{\tilde{p}_\perp} + s_q \frac{\tilde{q}_\|}{\tilde{q}_\perp} \right) &\longrightarrow& \tilde{p}_\| = \tilde{p}_\perp \frac{s_k \tilde{q}_\perp - s_q}{s_q \tilde{p}_\perp - s_p \tilde{q}_\perp}.
\end{eqnarray}
We obtain
\eq{energyIntegree}{
\begin{split}
    \pdt E_k =& \frac{\epsilon^2 C_E^2}{{2^{12}}\Omega_e} k_\perp^{4-2x} \left\vert k_\| \right\vert^{-2y} \sum_{s_k s_p s_q} \int_{\Delta_\perp} s_k s_p  \tilde{q}_\perp^{-x-y-2} \left(s_q\tilde{q}_\perp - s_p \tilde{p}_\perp  \right)^2 \left(s_k + s_p \tilde{p}_\perp + s_q \tilde{q}_\perp  \right)^2 \\
    &\times \sin \alpha_q \frac{s_k \tilde{q}_\perp - s_q}{s_q \tilde{p}_\perp - s_p \tilde{q}_\perp} \left\vert \frac{s_p - s_k \tilde{p}_\perp}{s_q \tilde{p}_\perp - s_p \tilde{q}_\perp} \right\vert^{-y} \left(1 - \tilde{p}_\perp^{-x-y-1} \left\vert \frac{s_k \tilde{q}_\perp - s_q}{s_q \tilde{p}_\perp - s_p \tilde{q}_\perp} \right\vert^{-y} \right) \\
    &\times \left\vert \frac{ \tilde{p}_\perp \tilde{q}_\perp }{s_p \tilde{q}_\perp - s_q \tilde{p}_\perp} \right\vert \dd \tilde{p}_\perp \dd \tilde{q}_\perp,
\end{split}
}
where $\sin \alpha_q= \sqrt{1 - \left(1 + \tilde{p}_\perp^2 - \tilde{q}_\perp^2 \right)^2 \left(2 \tilde{p}_\perp \right)^{-2}}$.

\subsection{Zone A}

We define $\tilde{p}_\perp = 1 + r \cos \beta$ and $\tilde{q}_\perp = r \sin \beta$, with $r \ll 1$ and $\beta \in \left[\pi/4, 3 \pi/4 \right]$. Two cases must be distinguished: when $s_k = s_p$ and when $s_k = -s_p$. An evaluation (to leading order) of the different terms of the integral (\ref{eq:energyIntegree}) is given in Table 1. Note that these evaluations take into account the possible cancellation of the integral due to $\beta$ symmetry. 
\begin{center}
\begin{tabular}{ |c|c|c| } 
 \hline
Table 1 & $s_k=s_p$ & $s_k=-s_p$ \\ 
 \hline
$\frac{s_k \tilde{q}_\perp - s_q}{s_q \tilde{p}_\perp - s_p \tilde{q}_\perp}$ & $ -1 $ & $ -1$ \\ 
$ \left\vert \frac{s_p - s_k \tilde{p}_\perp}{s_q \tilde{p}_\perp - s_p \tilde{q}_\perp} \right\vert $ & $ r \vert \cos \beta \vert $ & $ 2 $\\
$\left(s_q\tilde{q}_\perp - s_p \tilde{p}_\perp  \right)^2$ & 1 & 1 \\ 
$\left(s_k + s_p \tilde{p}_\perp + s_q \tilde{q}_\perp  \right)^2$ & 4 & $ r^2$ \\ 
$\sin \theta$ &  $r \sqrt{- \cos 2 \beta}$ & $r \sqrt{- \cos 2 \beta}$ \\
$ 1 - \tilde{p}_\perp^{-x-y-1} \left\vert \frac{s_k \tilde{q}_\perp - s_q}{s_q \tilde{p}_\perp - s_p \tilde{q}_\perp} \right\vert^{-y} $ & $ \propto r^2 \cos^2 \beta$ & $ \propto r^2 \cos^2 \beta$ \\ 
$\left\vert \frac{ \tilde{p}_\perp \tilde{q}_\perp }{s_p \tilde{q}_\perp - s_q \tilde{p}_\perp} \right\vert $ & $r \left\vert \sin \beta \right\vert$ & $r \left\vert \sin \beta \right\vert$ \\ 
$\dd \tilde{p}_\perp \dd \tilde{q}_\perp  $ & $r \dd r \dd \beta $ & $r \dd r \dd \beta $ \\ 
 \hline
\end{tabular}
\end{center}
When $s_k=s_p$, the criterion of convergence of the kinetic equation (\ref{eq:energyIntegree}) will be given by the following integral 
\eq{}{
    \int_0^{R<1} r^{3-x-2y} \dd r \int_{\pi/4}^{3\pi/4} {\left\vert \cos \beta \right\vert^{2-y}} \sqrt{- \cos 2 \beta} \left( \sin \beta \right)^{-1-x-y} \dd \beta .
}
Therefore, there is convergence if $x+2y < 4$.
When $s_k=-s_p$, we have
\eq{}{
    \int_0^{R<1} r^{5-x-y} \dd r \int_{\pi/4}^{3\pi/4} \cos^2 \beta \sqrt{-\cos 2 \beta} \left( \sin \beta \right)^{-1-x-y}  \dd \beta
}
and the convergence is obtained if $x + y < 6$.

\subsection{Zone B}

We define $\tilde{p}_\perp = r \cos \beta$ and $\tilde{q}_\perp = 1+ r \sin \beta$, with this time $\beta \in \left[-\pi/4, \pi/4 \right]$. We have two cases: $s_k=s_q$ and $s_k=-s_q$. An evaluation (to leading order) of the different terms of the integral (\ref{eq:energyIntegree}) is given in Table 2. Note that these evaluations take into account the possible cancellation of the integral due to $\beta$ symmetry. 
\begin{center}
\begin{tabular}{ |c|c|c| } 
 \hline
Table 2 & $s_k=s_q$ & $s_k=-s_q$ \\ 
 \hline
 $\frac{s_k \tilde{q}_\perp - s_q}{s_q \tilde{p}_\perp - s_p \tilde{q}_\perp}$ & $ s_k s_p r^2 \sin^2 \beta $ & $ - 2 s_k s_p $ \\
 $ \left\vert \frac{s_p - s_k \tilde{p}_\perp}{s_q \tilde{p}_\perp - s_p \tilde{q}_\perp} \right\vert$ & $ 1 $ & $ 1 $ \\
$\left(s_q\tilde{q}_\perp - s_p \tilde{p}_\perp  \right)^2$ & 1 & 1 \\ 
$\left(s_k + s_p \tilde{p}_\perp + s_q \tilde{q}_\perp  \right)^2$ & 4 & $ r^2$ \\ 
$\sin \theta$ &  $\sqrt{1 - \tan^2 \beta}$ & $\sqrt{1 - \tan^2 \beta}$ \\
$ 1 - \tilde{p}_\perp^{-x-y-1} \left\vert \frac{s_k \tilde{q}_\perp - s_q}{s_q \tilde{p}_\perp - s_p \tilde{q}_\perp} \right\vert^{-y} $ & $ 1 - \left( r \cos \beta \right)^{-x-y-1} \left\vert r \sin \beta \right\vert^{-y}$ & $ 1 - 2^{-y} \left( r \cos \beta \right)^{-x-y-1} $ \\ 
$\left\vert \frac{ \tilde{p}_\perp \tilde{q}_\perp }{s_p \tilde{q}_\perp - s_q \tilde{p}_\perp} \right\vert $ & $r \left\vert \cos \beta \right\vert$ & $r \left\vert \cos \beta \right\vert$ \\ 
$\dd \tilde{p}_\perp \dd \tilde{q}_\perp  $ & $r \dd r \dd \beta $ & $r \dd r \dd \beta $ \\ 
 \hline
\end{tabular}
\end{center}
When $s_k=s_q$, the criterion of convergence of the kinetic equation  (\ref{eq:energyIntegree}) will be given by the following integral
\eq{}{
    \int_0^{R<1} r^{3-x-2y} \dd r \int_{-\pi/4}^{+\pi/4} \left( \cos \beta \right)^{-x-y} \left\vert \sin \beta  \right\vert^{2-y} \sqrt{1 - \tan^2 \beta} \dd \beta .
}
Therefore, there is convergence if $x+2y < 4$.
When $s_k=-s_q$, we have
\eq{}{
     \int_0^{R<1} r^{3-x-y} \dd r \int_{-\pi/4}^{+\pi/4} \left( \cos \beta \right)^{-x-y} \sqrt{1 - \tan^2 \beta} \dd \beta
}
and the convergence is obtained if $x + y < 4$.

\subsection{Zone C}

We define $\tilde{p}_\perp = (\tau_2-\tau_1)/2$ and $\tilde{q}_\perp = (\tau_1+\tau_2)/2$, with $-1 \le \tau_1 \le 1$ and $1 \ll \tau_2$. We have two cases: $s_p=s_q$ and $s_p=-s_q$. An evaluation (to leading order) of the different terms of the integral (\ref{eq:energyIntegree}) is given in Table 3. Note that these evaluations take into account the possible cancellation of the integral due to $\tau_1$ symmetry.
\begin{center}
\begin{tabular}{ |c|c|c| } 
 \hline
 Table 3 & $s_p=s_q$ & $s_p=-s_q$ \\ 
 \hline
 $\frac{s_k \tilde{q}_\perp - s_q}{s_q \tilde{p}_\perp - s_p \tilde{q}_\perp}$ & $ - s_k s_p / 2 $ & $ - s_k s_p / 2 $ \\
 $ \left\vert \frac{s_p - s_k \tilde{p}_\perp}{s_q \tilde{p}_\perp - s_p \tilde{q}_\perp} \right\vert $ & $ \left\vert \tau_2 \tau_1^{-1} \right\vert / 2$ & $ 1/2 $ \\
 $\left(s_q\tilde{q}_\perp - s_p \tilde{p}_\perp  \right)^2$ & $\tau_1^2$ & $\tau_2^2$ \\ 
$\left(s_k + s_p \tilde{p}_\perp + s_q \tilde{q}_\perp  \right)^2$ & $\tau_2^2$  & $ 1 + \tau_1^2$ \\ 
$\sin \theta$  & $\sqrt{1 - \tau_1^2}$ & $\sqrt{1 - \tau_1^2}$ \\
$ 1 - \tilde{p}_\perp^{-x-y-1} \left\vert \frac{s_k \tilde{q}_\perp - s_q}{s_q \tilde{p}_\perp - s_p \tilde{q}_\perp} \right\vert^{-y} $ & $1 - 2^{x+2y+1} \tau_2^{-x-2y-1}  \left\vert \tau_1 \right\vert^y$ & $ 1 - 2^{x+y-1} \tau_2^{-x-y-1}$ \\
$\left\vert \frac{ \tilde{p}_\perp \tilde{q}_\perp }{s_p \tilde{q}_\perp - s_q \tilde{p}_\perp} \right\vert $ & $ \tau_2^2 \left\vert \tau_1^{-1} \right\vert / 4 $ & $ \tau_2 / 4 $ \\
$\dd \tilde{p}_\perp \dd \tilde{q}_\perp  $ & $ \propto \dd \tau_1 \dd \tau_2 $ & $ \propto \dd \tau_1 \dd \tau_2 $ \\ 
 \hline
\end{tabular}
\end{center}
When $s_p=s_q$, the criterion of convergence of the kinetic equation  (\ref{eq:energyIntegree}) 
will be given by the following integral
\eq{}{
    \int_{-1}^{+1} \sqrt{1-\tau_1^2} \left\vert \tau_1 \right\vert^{y+1} \dd \tau_1 \int_{\tau > 1}^{+\infty} \tau_2^{-x-2y+2} \dd \tau_2 .
}
Therefore, there is convergence if $3 < x + 2y$.
When $s_p=-s_q$, we have
\eq{}{
    \int_{-1}^{+1} \left( 1 + \tau_1^2 \right) \sqrt{1-\tau_1^2} \dd \tau_1 \int_{\tau>1}^{+\infty} \tau_2^{-x-y+1} \dd \tau_2,
}
and the convergence is obtained if $2 < x + y$.

In conclusion, a solution is local if the following conditions are satisfied
\begin{eqnarray}
3 < x &+& 2 y < 4 ,\\
2 < x &+& y < 4 .
\end{eqnarray}
We notice that the KZ spectrum for the energy corresponds to $x+2y=3.5$ and $x+y=3$. These values are thus exactly in the middle of the convergence domain. 

\bibliographystyle{jpp}
\bibliography{jpp-biblio}

\end{document}